\documentclass[twocolumn]{aastex701}

\usepackage{amsmath}
\usepackage{amsfonts}
\usepackage{graphicx}
\usepackage{rotating}
\usepackage{natbib}
\usepackage{txfonts}
\usepackage{color}
\usepackage{wrapfig}
\usepackage{amssymb}
\usepackage{color}
\usepackage{colortbl}
\usepackage{mhchem}

\shorttitle{Hot cores in the SNR RX~J1713.7$-$3946} 
\shortauthors{T. Shimonishi et al.} 

\begin{document}

\title{Survival of Molecular Complexity under Recent Supernova Feedback: Detection of Hot Cores in RX~J1713.7$-$3946}

\author[orcid=0000-0002-0095-3624]{Takashi Shimonishi}
\affiliation{Institute of Science and Technology, Niigata University, Ikarashi-ninocho 8050, Nishi-ku, Niigata 950-2181, Japan}
\email{shimonishi@env.sc.niigata-u.ac.jp}  
\correspondingauthor{Takashi Shimonishi} 
\email{shimonishi@env.sc.niigata-u.ac.jp} 

\author[orcid=0000-0003-2062-5692]{Hidetoshi Sano}
\affiliation{Department of Intelligence Science and Engineering, Graduate School of Natural Science and Technology, Gifu University, 1-1 Yanagido, Gifu, 501-1193 Japan}
\affiliation{Center for Space Research and Utilization Promotion (c-SRUP), Gifu University, 1-1 Yanagido, Gifu 501-1193, Japan}
\email{sano.hidetoshi.w4@f.gifu-u.ac.jp} 

\author[orcid=0000-0002-2026-8157]{Kenji Furuya}
\affiliation{RIKEN Pioneering Research Institute, 2-1 Hirosawa, Wako-shi, Saitama 351-0198, Japan}
\email{kenji.furuya@riken.jp} 

\author[orcid=0000-0002-0197-8751]{Yoko Oya}
\affiliation{Center for Gravitational Physics, Yukawa Institute for Theoretical Physics, Kyoto University, Oiwake-cho, Kitashirakawa, Sakyo-ku, Kyoto-shi, Kyoto-fu 606-0967, Japan}
\email{yoko.oya@yukawa.kyoto-u.ac.jp}

\begin{abstract}  
Protostellar cores located near supernova remnants are considered potential analogues of the birth environment of the solar system. 
However, the extent to which supernovae influence their chemical evolution remains unclear. 
We report the first detection of hot molecular cores in a supernova remnant using the Atacama Large Millimeter/submillimeter Array. 
The detected hot cores (HC1 and HC2) are located inside the X-ray shell of the young supernova remnant RX~J1713.7$-$3946, and both sources are associated with Class~I intermediate-mass protostars. 
This paper focuses on a detailed chemical analysis of HC1, in which a variety of carbon-, oxygen-, nitrogen-, sulfur-, and silicon-bearing species are detected. 
Excitation analyses indicate that HC1 harbors dense ($\sim$10$^{7}$~cm$^{-3}$), compact ($<$500~au), and high-temperature ($\gtrsim$100~K) molecular gas. 
Despite being located within a supernova-feedback region, the column density ratios of complex organic molecules (HCOOCH$_3$/CH$_3$OH, CH$_3$OCH$_3$/CH$_3$OH, and CH$_3$CHO/CH$_3$OH), a deuterated molecule (CH$_2$DOH/CH$_3$OH), and sulfur- and nitrogen-bearing species (OCS/CH$_3$OH and C$_2$H$_5$CN/CH$_3$CN) in HC1 are indistinguishable from those observed in hot cores/corinos in more typical star-forming environments. 
HC1 is located near the outer edge of the supernova shell, and the surrounding region has likely begun to be exposed to such a harsh environment only recently. 
The elapsed time since the onset of exposure to high-energy particles and photons may be too short for the chemical composition of the hot core to be significantly altered, and/or the hot-core region may be shielded by magnetic fields amplified by supernova feedback, which could suppress the penetration of enhanced cosmic rays. 
\end{abstract}


\section{Introduction} \label{sec_intro} 
\setcounter{footnote}{0}
Supernova explosions are among the most energetic phenomena that have a profound impact on the interstellar medium (ISM). 
Supernova explosions interact with the surrounding ISM through 
(i) intense cosmic-ray and X-ray radiation that is at least hundreds of times stronger than in typical interstellar environments,
(ii) powerful shock waves (several thousand km s$^{-1}$), which are at least an order of magnitude stronger than those generated by protostellar outflows, and 
(iii) the injection of rare elements (e.g., phosphorus) and rare isotopic species \citep[e.g.,][and references therein]{Koo13, Sla15, 2021Ap&SS.366...58S}. 
Such supernova feedback is expected to have a significant impact on the chemical evolution of the ISM. 

The analysis of short-lived radioisotopes in meteorites and asteroid samples has revealed the presence of distinct patterns in the isotopic compositions of primitive solar system materials \citep[e.g.,][]{Lee76, Sco07}. 
A leading hypothesis to explain these isotopic anomalies is that the protosolar system was influenced by supernova-driven processes, such as the injection of material or cosmic-ray nucleosynthesis in a shockwave \citep[e.g.,][and references therein]{Ada10, Bos12, Saw25}. 
Therefore, protostellar cores located near supernova remnants (SNRs) are of great interest as potential analogues of the birth environment of the solar system.

The chemistry of complex organic molecules (COMs) is one of the key issues in astrochemistry. 
Abundant gaseous COMs are often detected in high-temperature ($>$100 K), high-density ($\gtrsim$10$^6$ cm$^{-3}$), and compact ($\ll$0.1 pc) protostellar envelopes \citep[e.g.,][]{Her09}. 
\added{The majority of COMs are believed to be initially produced in the ice mantles of dust grains and subsequently released into the gas phase through thermal desorption as the grains are heated by star formation activity \citep[e.g.,][]{Gar06,vanD25}. 
In addition, neutral-neutral reactions in high-temperature gas may further enhance molecular complexity following the desorption of ice-mantle species \citep[e.g.,][and references therein]{Oya16,Cec23}. 
Such ice-sublimation regions associated with protostars across a wide range of masses are referred to as hot cores \citep[e.g.,][]{Mor80,Kur00,Sch02}, whereas their low-mass counterparts are sometimes called hot corinos \citep[e.g.,][]{Cec04}. }

The relationship between high-energy particles/photons and the emergence of chemical complexity has been attracting increasing attention \citep[e.g.,][]{Aru19,Gac25}.
However, the net impact of energetic processing on the chemical complexity of star-forming material remains unclear, as it may either enhance or suppress molecular abundances depending on the evolutionary stage at which the material is exposed to energetic particles and radiation. 
Energetic processing during the prestellar phase may promote the formation of COMs in ice mantles. 
Laboratory studies have shown that irradiation of interstellar ice analogues by energetic particles or electrons contributes to the formation of organic molecules by generating reactive molecular radicals that subsequently form COMs via radical–radical reactions \citep[e.g.,][]{Kai98,Hud99,Ben07}. 
Such processes may be enhanced in environments exposed to intense cosmic rays and X-rays. 
In contrast, once ice mantles sublimate in the protostellar phase, excessive cosmic rays, X-rays, and the resultant secondary ultraviolet photons may instead lead to the destruction of gas-phase molecules and reduce their abundances. 
Observational approaches are essential for understanding how energetic particles and radiation influence the chemical evolution of star- and planet-forming regions, as it remains challenging for astrochemical simulations to account for the complex interplay of energetic and non-thermal chemical processes. 

Single-dish observations of molecular clouds around SNRs suggest that supernova feedback (particularly intense cosmic rays) leads to the efficient dissociation of CO and enhanced production of molecular ions such as \ce{HCO+} \citep[e.g.,][]{Yam23,Tu24}. 
\added{Detections of cold molecular gas species, including several COMs (e.g., CH$_3$OH, CH$_3$CHO, and CH$_3$CN), in dense clumps near SNRs have also been reported by single-dish observations \citep{Max12,Cos26}. }
\added{In other environments with enhanced cosmic rays, such as the Galactic Center region, various COMs have been detected toward cold, dense, shocked molecular clouds \citep[e.g.,][and references therein]{Zen18,Jim25}. }
Nevertheless, the effects of supernova feedback on the chemical complexity of protostellar cores remain poorly understood due to the limited availability of high-spatial-resolution observational data. 
Chemical analyses of compact, dense, and high-temperature regions associated with protostars are essential for understanding the impact of supernova feedback on the chemical evolution of COMs in star- and planet-forming regions. 

We here report the first detection of hot molecular cores in a supernova-feedback region\footnote{In this study, the region inside the X-ray shell is defined as the supernova-feedback region.} based on submillimeter observations with the Atacama Large Millimeter/submillimeter Array (ALMA). 
This study represents the first observational attempt to characterize the chemical evolution of protostellar cores located within a SNR.

\begin{deluxetable*}{ l c c c c c c c c c}
\tablecaption{Observation summary \label{tab_Obs}} 
\tablewidth{0pt} 
\tabletypesize{\footnotesize} 
\tablehead{
\colhead{}   & \colhead{Observation} &  \colhead{On-source}               & \colhead{Mean}                             & \colhead{Number}   &  \multicolumn{2}{c}{Baseline}     &  \colhead{}                                                  &  \colhead{}                                         &  \colhead{Channel}  \\
\cline{6-7}  
\colhead{}   & \colhead{date}             &  \colhead{time per target}         &  \colhead{PWV\tablenotemark{a}} & \colhead{of}              &  \colhead{Min} & \colhead{Max} & \colhead{Beam size\tablenotemark{b}}        & \colhead{MRS\tablenotemark{c}}     &  \colhead{spacing}  \\
\colhead{}   & \colhead{}                    &  \colhead{(min)}                        &  \colhead{(mm)}                             & \colhead{antennas} & \colhead{(m)} & \colhead{(m)}   &  \colhead{($\arcsec$ $\times$ $\arcsec$)} &  \colhead{($\arcsec$)}                       &  \colhead{}                }
\startdata
Band 6       &  2025 Jan 9             &  65                          &  1.3--1.9                                         &  45--49                     &  15.1               &  783.5              &  0.39 $\times$ 0.47                                     &   5.7                              &  0.49 MHz (0.58 km s$^{-1}$)              \\
Band 7       &  2024 Dec 26--27    &  75                           &  0.6--1.0                                         &  38-43                       &  15.0               &  499.8              &  0.42 $\times$ 0.53                                     &   5.1                              &  0.49 MHz (0.43 km s$^{-1}$)                \\
\enddata
\tablenotetext{a}{Precipitable water vapor.}
\tablenotetext{b}{The average synthesized beam size of spectral cube data. }
\tablenotetext{c}{Maximum Recoverable Scale.}
\end{deluxetable*}

\begin{figure*}[ptb] 
\begin{center} 
\includegraphics[width=18.0cm]{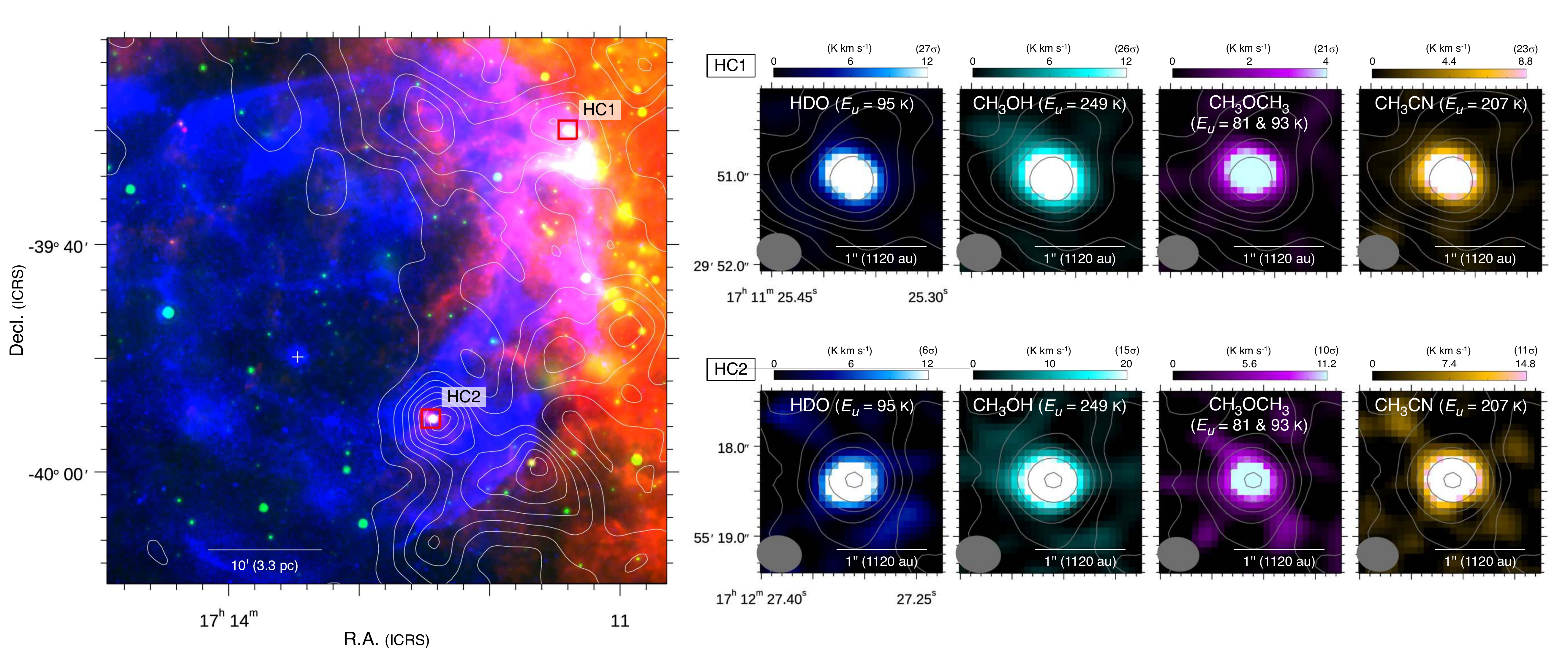} 
\caption{
Left: Three-color composite image of the supernova remnant RX~J1713.7$-$3946 (red: Herschel/PACS 160 $\mu$m; green: WISE 22 $\mu$m; blue: XMM-Newton X-ray) \citep[][DOI: 10.26131/IRSA79 and 10.26131/IRSA535]{Pib10, Wri10, 2021ApJ...915...84F}. 
The white contours show the total proton column density (H$_2$ + H) estimated from CO and H\,\textsc{i} observations \citep{Fuk12}. 
The contour levels are 6, 7, 8, 10, 12, 14, 16, 18, 20, and 22 $\times$ 10$^{21}$ cm$^{-2}$. 
The red squares indicate the locations of the two hot cores (RX1713 HC1 and HC2) discovered in this work. 
The cross indicates the position of the neutron star (1WGA J1713.4-3949), a possible progenitor of RX1713 \citep{Laz03}. 
Right: ALMA integrated intensity images of the selected high-excitation lines associated with HC1 and HC2. 
HDO(2$_{1,1}$--2$_{1,2}$), CH$_3$OH(14$_{-1}$ E--13$_{-2}$ E), CH$_3$OCH$_3$(13$_{1,13}$--12$_{0,12}$ EE and 14$_{1,14}$--13$_{0,13}$ EE, stacked), CH$_3$CN(14$_{4}$--13$_{4}$) are plotted. 
\added{The noise level corresponding to the top of the color bar (white) is indicated in parentheses. }
Contours represent the 1.2~mm continuum and the contour levels are 
10$^{0.8}$, 10$^{1.1}$, 10$^{1.4}$, 10$^{1.7}$, 10$^{2.0}$, 10$^{2.3}$, and 10$^{2.6}$ $\sigma$, where $\sigma$ (image rms) is 0.02 mJy beam$^{-1}$. 
The synthesized beam is shown as a gray filled ellipse. 
Images for all detected emission lines are presented in Appendix~\ref{sec_app_img}. 
}
\label{rx1713}
\end{center}
\end{figure*}
%

\section{Target, observations, and data reduction} \label{sec_tarobsred} 
\subsection{Target} \label{sec_tar}
The target supernova remnant is RX~J1713.7$-$3946 (hereafter, RX1713), which is one of the high-energy $\gamma$-ray core-collapse SNRs associated with dense molecular clouds \citep[e.g.,][]{San10}. 
It is located at the distance of 1.12 $\pm$ 0.01 kpc, as precisely estimated by a dust extinction study of nearby GAIA sources \citep{Lei21}. 
RX1713 is a relatively young SNR with an estimated age of approximately 1600~yr \citep{Wan97}. 
\added{The cosmic-ray ionization rate in this region is expected to be $\sim$10–100 times higher than the standard value in molecular clouds, reaching $\sim$10$^{-15}$~$\mathrm{s^{-1}}$, according to measurements toward similar SNRs \citep[e.g.,][]{Ind10,Cec11}. }
X-ray observations have reported that RX1713 is associated with shock waves of $\sim$1000--4000 km s$^{-1}$ \citep{Tsu16}. 
These previous studies indicate that the dense clouds associated with RX1713 have been exposed to harsh and energetic environments not typically found in commonly studied nearby star-forming regions. 

We conducted ALMA observations toward two fields in the RX1713 region. 
These two fields correspond to Peak D and Peak C of the CO(2-1) emission reported by \citet{San10}. 
The Peak D field includes a source that exhibits signatures of a bipolar molecular outflow traced by CO gas in the ALMA archival data (ID: 2017.1.01406.S). 
In the Peak C field, signatures of a bipolar outflow have been identified based on low-spatial-resolution CO observations with NANTEN2 \citep{San10}. 
The presence of protostellar outflows is a good indicator of ongoing star formation. 
The locations of the two target fields within the RX1713 region are shown in Figure~\ref{rx1713}.

\subsection{Observations} \label{sec_obs} 
Observations were conducted with ALMA in 2024 and 2025 as a part of the Cycle 11 program (2024.1.00402.S, PI: T. Shimonishi). 
Eight spectral windows are used in total to cover the rest frequencies of 241.46--243.34, 243.82-245.69, 256.96--258.83, 258.82--260.69 GHz in Band 6, and 337.28--339.16, 339.05-340.93, 349.17--351.04, and 350.97--352.84 GHz in Band 7. 
Details of the observation settings are summarized in Table~\ref{tab_Obs}.

\subsection{Data reduction} \label{sec_red} 
Raw data was processed with the \textit{Common Astronomy Software Applications} (CASA) package \citep[version 6.6.1,][]{casa2022}. 
The pipeline-calibrated visibility data provided by the observatory was used for the imaging process. 
The CASA task \texttt{tclean} was used for imaging and the masking was done using the auto-multithresh algorithm. 
The synthesized beam size is 
0$\farcs$39--0$\farcs$41 $\times$ 0$\farcs$47--0$\farcs$50 for Band 6 and 
0$\farcs$42--0$\farcs$43 $\times$ 0$\farcs$52--0$\farcs$54 for Band 7 
with the Briggs weighting and a robustness parameter of 0.5. 
The synthesized images were corrected for the primary beam pattern using the \texttt{impbcor} task in CASA. 
The continuum image was constructed by selecting line-free channels. 
The continuum emission was subtracted from the spectral data using the \texttt{uvcontsub} task in CASA before their processing.

\begin{figure*}[tp!] 
\begin{center} 
\includegraphics[width=17.5cm]{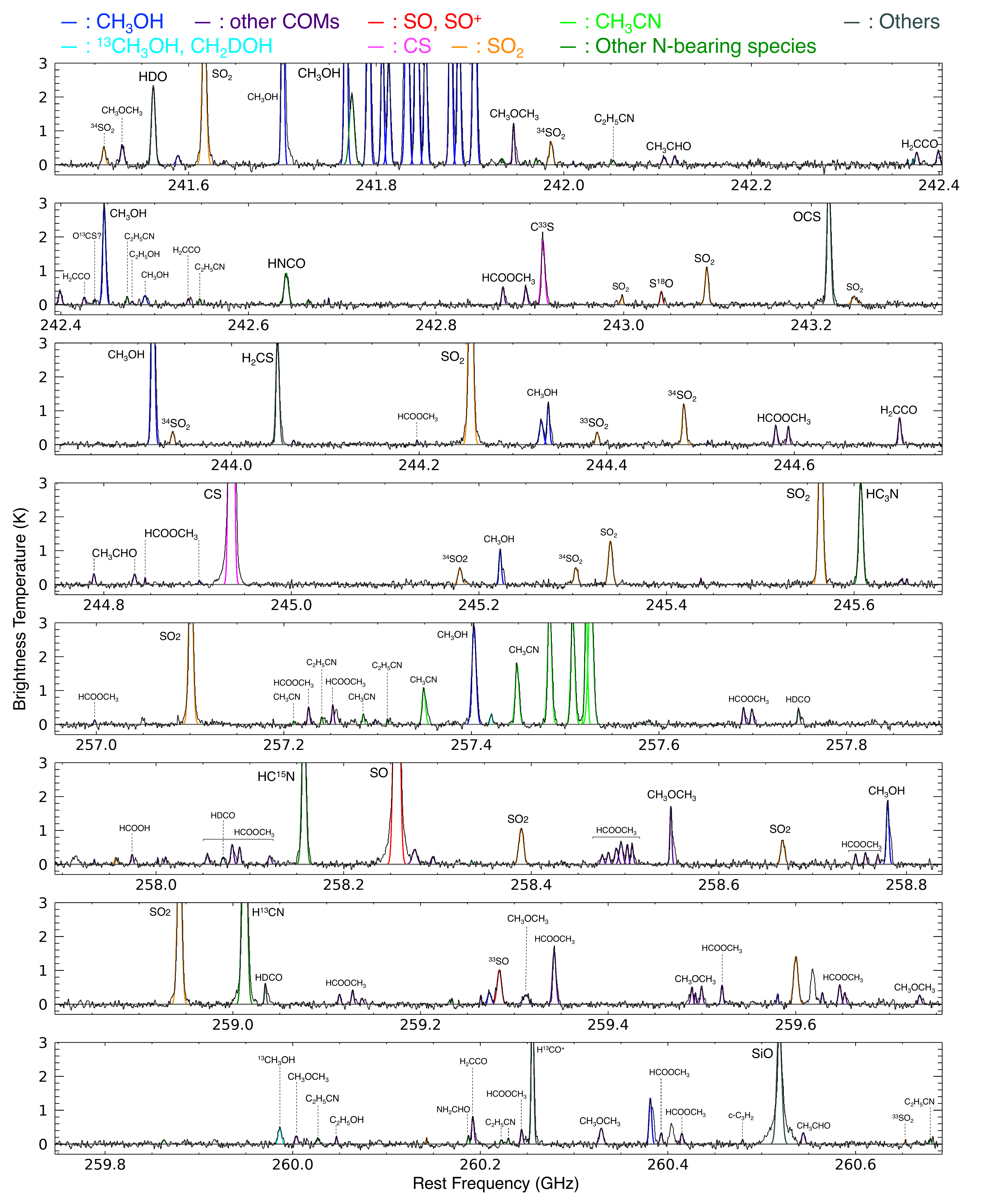} 
\caption{
ALMA Band 6 spectra of RX1713 HC1. 
The black line represents the observed spectra, while the colored lines indicate the line fitting results. 
Detected emission lines are labeled. 
Tentative detections are indicated by ``?". 
The source velocity of $-$8.0 km s$^{-1}$ is assumed. 
}
\label{spec_B6}
\end{center}
\end{figure*}
%
\begin{figure*}[tp!] 
\begin{center} 
\includegraphics[width=17.5cm]{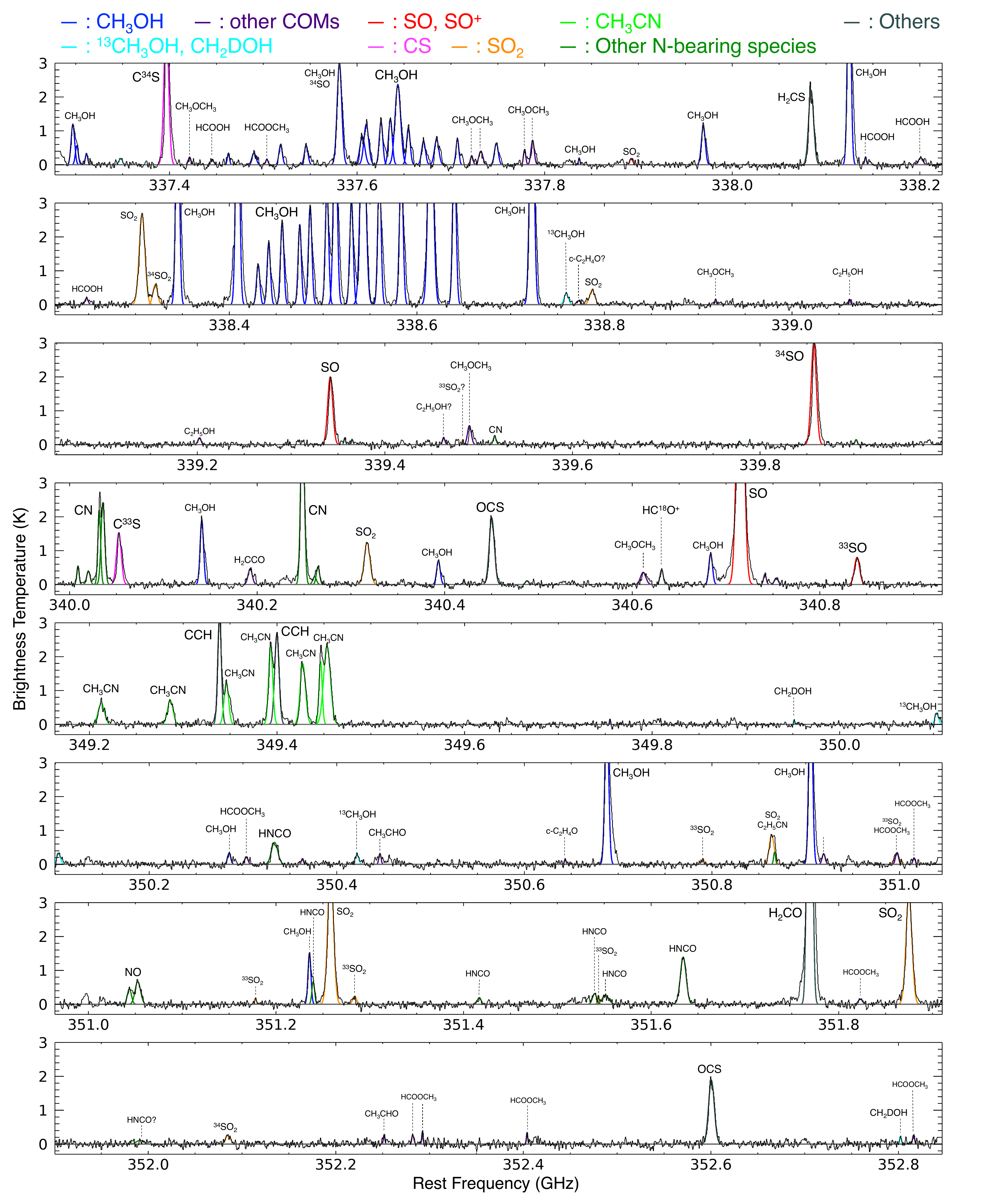} 
\caption{
Same as in Figure \ref{spec_B6}, but for Band 7. 
}
\label{spec_B7}
\end{center}
\end{figure*}

\begin{deluxetable*}{ l l l l l l l l }[tp!]
\tablecaption{Summary of detected molecular species \label{tab_line_summary}}
\tablewidth{0pt}
\tabletypesize{\footnotesize} 
\tablehead{
\colhead{Two atoms}   & \colhead{Three atoms}   &       \colhead{Four atoms}   &    \colhead{Five atoms}&       \colhead{Six atoms}  &       \colhead{Seven atoms} &       \colhead{Eight atoms} &       \colhead{Nine atoms}            \\
}
\startdata 
CN                          &   HDO                       &      H$_2$CO               &         HCOOH              &           CH$_3$OH       &         CH$_3$CHO      &  HCOOCH$_3$          &   CH$_3$OCH$_3$          \\
NO                          &   H$^{13}$CO$^+$   &      HDCO                     &         H$_2$CCO         &     $^{13}$CH$_3$OH  &    (c-C$_2$H$_4$O)   &                                   &    C$_2$H$_5$OH                \\
CS                          &    HC$^{18}$O$^+$  &       HNCO                     &        HC$_3$N             &         CH$_2$DOH      &                                    &                                   &   C$_2$H$_5$CN          \\
C$^{34}$S              &   H$^{13}$CN           &       H$_2$CS                &        (c-C$_3$H$_2$)  &          CH$_3$CN         &                                   &                                    &       \\
C$^{33}$S              &   HC$^{15}$N           &                                      &                                     &  ($^{13}$CH$_3$CN)   &                                    &                                    &       \\
SO                          &   CCH                       &                                        &                                     &      NH$_2$CHO      &                                      &                                    &       \\
$^{34}$SO              &  SO$_2$                   &                                      &                                    &                                    &                                      &                                    &       \\
$^{33}$SO              &   $^{34}$SO$_2$      &                                     &                                    &                                    &                                      &                                    &       \\
S$^{18}$O                  &   $^{33}$SO$_2$      &                                     &                                     &                                    &                                      &                                    &       \\
SiO                         &   OCS                       &                                     &                                      &                                    &                                      &                                    &       \\
                               &   O$^{13}$CS         &                                     &                                      &                                    &                                      &                                    &       \\
                               &   (HDS)                     &                                     &                                      &                                    &                                      &                                    &       \\
\enddata
\tablecomments{
Molecular species enclosed in parentheses indicate tentative detections. 
}
\end{deluxetable*}

\section{Results and analysis} \label{sec_res} 
As a result of the observations, a compact point source associated with strong continuum emission and high-excitation molecular line emission is identified in each of the two observed fields (see Figure~\ref{rx1713}). 
Hereafter, we refer to these two sources as RX1713 HC1 and HC2. 
The detection of these emission, which trace high-temperature gas surrounding protostars, suggests that they are protostellar objects associated with hot molecular cores. 
\added{The systemic velocities of the molecular gas associated with HC1 and HC2 are about $-8.0$ km s$^{-1}$ and $-12.5$ km s$^{-1}$, respectively (Section~\ref{sec_spc}). 
\citet{San13} mapped the distribution of molecular clouds interacting with the SNR based on a comparison between X-ray and CO observations. 
The two sources identified in this work are consistent with the systemic velocities of these molecular clouds and are located near the peaks of their spatial distribution (see Clump D and C in their Fig.5). 
This indicates that both sources are protostellar objects physically associated with the SNR. }

The peak positions of the continuum and molecular line emission are nearly coincident, and the corresponding coordinates are as follows: 
for HC1, RA = 17$^{\mathrm h}$11$^{\mathrm m}$25$\fs$373 and Dec = $-39\arcdeg29\arcmin51\farcs00$ (ICRS); 
and for HC2, RA = 17$^{\mathrm h}$12$^{\mathrm m}$27$\fs$312 and Dec = $-39\arcdeg55\arcmin18\farcs35$. 
The position of HC1 coincides with that of the CO bipolar outflow source identified in the ALMA archival data. 
\added{In this paper, we focus on a detailed chemical analysis of HC1
\footnote{The other hot core, HC2, is located very close to the edge of the Band~6 field of view, and no Band~7 data are currently available. Follow-up ALMA observations are ongoing, and detailed chemical analyses will be presented in a forthcoming paper.}
. }

\subsection{Spectra} \label{sec_spc} 
The spectra and continuum flux of HC1 are extracted from a circular region with a diameter of 0$\farcs$50 (560 au at the distance of RX1713), centered at its continuum and molecular emission peak mentioned above. 
Upon spectral extraction, all images are convolved to have a common circular beam size of 0$\farcs$54. 

Figures \ref{spec_B6}--\ref{spec_B7} show the extracted ALMA spectra. 
Spectral lines are identified with the aid of the Cologne Database for Molecular Spectroscopy\footnote{https://cdms.astro.uni-koeln.de} \citep[CDMS,][]{Mul05} and the molecular database of the Jet Propulsion Laboratory\footnote{http://spec.jpl.nasa.gov} \citep[JPL,][]{Pic98}. 
The detection criteria adopted here are a 3$\sigma$ significance level and velocity coincidence with the systemic velocity ($V_{\rm sys}$) of HC1 ($-8.0$ km s$^{-1}$), corresponding to the typical $V_{\rm sys}$ of the CH$_3$OH lines detected in HC1. 
The lines with the significance level higher than 2.5$\sigma$ but lower than 3$\sigma$ are identified as tentative detection. 

A variety of carbon-, oxygen-, nitrogen-, sulfur-, and silicon-bearing species, including COMs containing up to nine atoms, are detected toward RX1713 HC1 (see Table~\ref{tab_line_summary})
Multiple high-excitation lines (upper-state energies $>100$ K) are detected for many species.
Line parameters are measured by fitting a Gaussian profile to detected lines, and the full details can be found in Appendix~\ref{sec_app_lineparam}. 
\added{The measured line widths are typically 4–6 km s$^{-1}$, comparable to those of known hot cores \citep[e.g.,][]{Bis07,Taq15,ST21}. 
For several molecular lines, wing-like components deviating from the Gaussian profile are observed (see Section~\ref{sec_disc_phys}). 
The velocity ranges corresponding to these wing components are excluded from the Gaussian fitting. }

\subsection{Images} \label{sec_img} 
The spatial distributions of the continuum and molecular line emission are shown in Appendix~\ref{sec_app_img}. 
Many of the detected molecular lines show intensity peaks at the continuum center, corresponding to the position of a protostar. 
CCH and CN exhibit peaks slightly offset from the hot-core position and display extended structures. 
H$^{13}$CO$^+$, HC$^{18}$O$^+$, H$^{13}$CN, CS, SO, and H$_2$CO show emission that extends beyond the hot core. 
For SiO and SO, elongated structures extending toward the northeast from the hot-core position are prominent \added{(see Section~\ref{sec_disc_phys} for further discussion). }

We also estimated the deconvolved source sizes as 
$\sqrt{\mathrm{FWHM_{maj}} \times \mathrm{FWHM_{min}} - \Theta_{\mathrm{beam}}^2 }$, 
where $\mathrm{FWHM_{maj}}$ and $\mathrm{FWHM_{min}}$ are the major and minor FWHM sizes derived from two-dimensional Gaussian fits, and $\Theta_{\mathrm{beam}}$ is the geometric mean of the beam major and minor axes (see Table~\ref{tab_N}). 
Note that only upper limits are estimated for emission whose deconvolved source sizes are smaller than half of the beam size.

\begin{deluxetable*}{ l c c c c c c }[p!]
\tablecaption{Estimated rotation temperatures, column densities, abundances, and deconvolved source sizes \label{tab_N}}
\tabletypesize{\scriptsize} 
\tablehead{
\colhead{Molecule}  & \colhead{$T_{\mathrm{rot}}$}  &  \colhead{$N$(X)}      & \colhead{$N_{non-LTE}$(X)} & \colhead{$N$(X)/$N_{\mathrm{H_2}}$} &  Size$_{\mathrm{deconv}}$ & Note  \\
\colhead{}          & \colhead{(K)}                 &  \colhead{(cm$^{-2}$)} & \colhead{(cm$^{-2}$)}      & \colhead{}                          & \colhead{(arcsec/au)}  & \colhead{} 
}
\startdata 
H$_2$                                     &  \nodata            &  (8.0 $\pm$ 2.0) $\times$ 10$^{22}$  &  \nodata                                            &  \nodata                                &  0.5/560        & Based on continuum, $T_{\mathrm{dust}}$ = 100--150 K assumed  \\
HCO$^+$ ($^{13}$C)                        &  \nodata            &  (6.1 $\pm$ 1.8) $\times$ 10$^{14}$  &  \nodata                                            &  (7.6 $\pm$ 3.0) $\times$ 10$^{-9}$     &  \nodata        & $N$ derived from $^{13}$C isotopologue  \\
H$^{13}$CO$^+$                            &  \nodata            &  (9.2 $\pm$ 0.5) $\times$ 10$^{12}$  &  (9.7 $\pm$ 0.3) $\times$ 10$^{12}$                 &  (1.2 $\pm$ 0.3) $\times$ 10$^{-10}$    &    1.1/1230     & $T_{\mathrm{rot}}$ = 20--40 K assumed  \\
HC$^{18}$O$^+$                            &  \nodata            &  (7.7 $\pm$ 2.6) $\times$ 10$^{11}$  &  (7.7 $\pm$ 1.7) $\times$ 10$^{11}$                 &  (9.6 $\pm$ 4.0) $\times$ 10$^{-12}$    &  1.1/1230       & $T_{\mathrm{rot}}$ = 20--40 K assumed  \\
C$_2$H                                    &  \nodata            &  (3.4 $\pm$ 0.8) $\times$ 10$^{14}$  &  \nodata                                            &  (4.2 $\pm$ 1.4) $\times$ 10$^{-9}$     &  $>$2/$>$2200   & $T_{\mathrm{rot}}$ = 20--40 K assumed  \\
c-C$_3$H$_2$$\dag$                        &  \nodata            &  (1.3 $\pm$ 0.9) $\times$ 10$^{13}$  & (1.2 $\pm$ 0.9) $\times$ 10$^{13}$\tablenotemark{a} &  (1.6 $\pm$ 1.2) $\times$ 10$^{-10}$    &  \nodata        & $T_{\mathrm{rot}}$ = 20--40 K assumed  \\
H$_2$CO                                   &  \nodata            &  (5.0 $\pm$ 2.1) $\times$ 10$^{14}$  & (4.1 $\pm$ 0.5) $\times$ 10$^{14}$\tablenotemark{a} &  (6.3 $\pm$ 3.0) $\times$ 10$^{-9}$     &  0.90/1010      & $T_{\mathrm{rot}}$ = 50--150 K assumed   \\
HDCO                                      &  \nodata            &  (4.5 $\pm$ 2.5) $\times$ 10$^{13}$  &  \nodata                                            &  (5.7 $\pm$ 3.4) $\times$ 10$^{-10}$    &  0.85/950       & $T_{\mathrm{rot}}$ = 50--150 K assumed  \\
D$_2$CO                                   &  \nodata            &  $<$3 $\times$ 10$^{12}$             &  \nodata                                            &  $<$4 $\times$ 10$^{-11}$               &  \nodata        & $T_{\mathrm{rot}}$ = 50--150 K assumed  \\
CH$_3$OH                                  &  143$^{+1}_{-1}$    &  1.7 $\times$ 10$^{16}$              & 1--9 $\times$ 10$^{16}$\tablenotemark{b}            & \nodata                                 &  0.25/280       & Likely optically thick  \\
CH$_3$OH ($^{13}$C)                       &  \nodata            &  (5.4 $\pm$ 1.7) $\times$ 10$^{16}$  &  \nodata                                            &   (6.7 $\pm$ 2.7) $\times$ 10$^{-7}$    &  \nodata        & $N$ derived from $^{13}$C isotopologue  \\
$^{13}$CH$_3$OH                           &  135$^{+14}_{-12}$  &  (8.1 $\pm$ 1.0) $\times$ 10$^{14}$  &  \nodata                                            &   (1.0 $\pm$ 0.3) $\times$ 10$^{-8}$    &  0.41/460       &  \\
CH$_2$DOH                                 &  \nodata            &  (2.7 $\pm$ 1.1) $\times$ 10$^{14}$  &  \nodata                                            &   (3.3 $\pm$ 1.6) $\times$ 10$^{-9}$    &  \nodata        & $T_{\mathrm{rot}}$ = $T_{\mathrm{rot}}$($^{13}$CH$_3$OH) assumed \\
trans-HCOOH                               &  190$^{+79}_{-43}$  &  (1.5 $\pm$ 0.5) $\times$ 10$^{14}$  &  \nodata                                            &   (1.9 $\pm$ 0.8) $\times$ 10$^{-9}$    &  0.41/460       &  \\
cis-HCOOH$\dag$                           &  \nodata            &  (2.0 $\pm$ 1.0) $\times$ 10$^{13}$  &  \nodata                                            &   (2.5 $\pm$ 1.4) $\times$ 10$^{-10}$   &  \nodata        & $T_{\mathrm{rot}}$ = $T_{\mathrm{rot}}$(trans-HCOOH) assumed  \\
CH$_3$CHO                                 &  52$^{+4}_{-4}$     &  (1.8 $\pm$ 0.3) $\times$ 10$^{14}$  &  \nodata                                            &   (2.3 $\pm$ 0.7) $\times$ 10$^{-9}$    &  0.64/720       &  \\
CH$_3$OCH$_3$                             &  102$^{+4}_{-4}$    &  (1.5 $\pm$ 0.1) $\times$ 10$^{15}$  &  \nodata                                            &   (1.8 $\pm$ 0.5) $\times$ 10$^{-8}$    &  0.26/290       &  \\
C$_2$H$_5$OH                              &  \nodata            &  (4.3 $\pm$ 1.2) $\times$ 10$^{14}$  &  \nodata                                            &   (5.4 $\pm$ 2.1) $\times$ 10$^{-9}$    &  $<$0.21/$<$240 & $T_{\mathrm{rot}}$ = $T_{\mathrm{rot}}$(CH$_3$OCH$_3$) assumed \\
HCOOCH$_3$                                &  108$^{+3}_{-3}$    &  (1.4 $\pm$ 0.1) $\times$ 10$^{15}$  &  \nodata                                            &   (1.8 $\pm$ 0.5) $\times$ 10$^{-8}$    &  $<$0.21/$<$240 &  \\
H$_2$CCO                                  &  74$^{+7}_{-6}$     &  (1.5 $\pm$ 0.2) $\times$ 10$^{14}$  &  \nodata                                            &   (1.8 $\pm$ 0.5) $\times$ 10$^{-9}$    &  0.33/370       &  \\
c-C$_2$H$_4$O$\dag$                       &  \nodata            &  (2.0 $\pm$ 1.0) $\times$ 10$^{13}$  &  \nodata                                            &   (2.5 $\pm$ 1.4) $\times$ 10$^{-10}$   &  \nodata        & $T_{\mathrm{rot}}$ = $T_{\mathrm{rot}}$($^{13}$CH$_3$OH) assumed  \\
HDO                                       &  164$^{+30}_{-22}$  &  (3.0 $\pm$ 0.3) $\times$ 10$^{15}$  &  \nodata                                            &  (3.7 $\pm$ 1.0) $\times$ 10$^{-8}$     &  $<$0.23/$<$260 &  \\
CN                                        &  \nodata            &  (1.8 $\pm$ 0.5) $\times$ 10$^{14}$  & (1.8 $\pm$ 0.4) $\times$ 10$^{14}$                  &   (2.3 $\pm$ 0.8) $\times$ 10$^{-9}$    &  $>$2/$>$2200   & $T_{\mathrm{rot}}$ = 20--40 K assumed  \\
HCN ($^{13}$C)                            &  \nodata            &  (1.0 $\pm$ 0.3) $\times$ 10$^{16}$  &  \nodata                                            &   (1.3 $\pm$ 0.5) $\times$ 10$^{-7}$    &  \nodata        & $N$ derived from $^{13}$C isotopologue  \\
H$^{13}$CN                                &  \nodata            &  (5.1 $\pm$ 0.2) $\times$ 10$^{13}$  & (6.6 $\pm$ 0.2) $\times$ 10$^{13}$                  &   (6.4 $\pm$ 1.6) $\times$ 10$^{-10}$   &  0.44/490       & $T_{\mathrm{rot}}$ = 20--40 K assumed  \\
HC$^{15}$N                                &  \nodata            &  (2.5 $\pm$ 0.1) $\times$ 10$^{13}$  & (2.6 $\pm$ 0.1) $\times$ 10$^{13}$                  &   (3.1 $\pm$ 0.8) $\times$ 10$^{-10}$   &  0.33/370       & $T_{\mathrm{rot}}$ = 20--40 K assumed  \\
HNCO                                      &  103$^{+4}_{-4}$    &  (2.6 $\pm$ 0.3) $\times$ 10$^{14}$  &  \nodata                                            &   (3.3 $\pm$ 0.9) $\times$ 10$^{-9}$    &  0.51/570       &  \\
NO                                        &  \nodata            &  (5.2 $\pm$ 0.6) $\times$ 10$^{15}$  &  (4.0 $\pm$ 0.8) $\times$ 10$^{15}$                 &   (6.5 $\pm$ 1.8) $\times$ 10$^{-8}$    &  0.59/660       & $T_{\mathrm{rot}}$ = $T_{\mathrm{rot}}$(HNCO) assumed  \\
CH$_3$CN                                  &  111$^{+2}_{-2}$    &  (2.0 $\pm$ 0.1) $\times$ 10$^{14}$  &  (2.1 $\pm$ 0.1) $\times$ 10$^{14}$                 &   (2.5 $\pm$ 0.6) $\times$ 10$^{-9}$    &  0.24/270       &  \\
$^{13}$CH$_3$CN$\dag$                     &  \nodata            &  (2.9 $\pm$ 1.3) $\times$ 10$^{12}$  &  \nodata                                            &   (3.6 $\pm$ 1.8) $\times$ 10$^{-11}$   &  \nodata        & $T_{\mathrm{rot}}$ = $T_{\mathrm{rot}}$(CH$_3$CN) assumed  \\
HCCN                                      &  \nodata            &  $<$1 $\times$ 10$^{12}$             &  \nodata                                            &  $<$1 $\times$ 10$^{-11}$               &   \nodata       & $T_{\mathrm{rot}}$ = $T_{\mathrm{rot}}$(CH$_3$CN) assumed  \\
HC$_3$N                                   &  \nodata            &  (8.1 $\pm$ 0.2) $\times$ 10$^{13}$  &  (7.9 $\pm$ 0.2) $\times$ 10$^{13}$                 &  (1.0 $\pm$ 0.3) $\times$ 10$^{-9}$     &  0.33/370       & $T_{\mathrm{rot}}$ = $T_{\mathrm{rot}}$(CH$_3$CN) assumed  \\
C$_2$H$_5$CN                              &  155$^{+193}_{-55}$ &  (3.8 $\pm$ 2.2) $\times$ 10$^{13}$  &  \nodata                                            &  (4.7 $\pm$ 3.0) $\times$ 10$^{-10}$    &  $<$0.22/$<$250 &  \\
NH$_2$CHO                                 &  102$^{+31}_{-19}$  &  (1.2 $\pm$ 0.7) $\times$ 10$^{13}$  &  \nodata                                            &  (1.5 $\pm$ 1.0) $\times$ 10$^{-10}$    &  $<$0.24/$<$270 &  \\
SiO                                       &  \nodata            &  (6.5 $\pm$ 1.8) $\times$ 10$^{13}$  &  (3.4 $\pm$ 0.2) $\times$ 10$^{13}$                 &  (8.1 $\pm$ 3.0) $\times$ 10$^{-10}$    &  $>$1/$>$1120   & $T_{\mathrm{rot}}$ = 50--150 K assumed, elongated distribution  \\
CS                                        &  \nodata            &  4.2 $\times$ 10$^{14}$              &  (1.3 $\pm$ 0.1) $\times$ 10$^{15}$                 &  \nodata                                &  0.93/1040      & Likely optically thick  \\
CS ($^{34}$S)                             &  \nodata            &  (2.0 $\pm$ 0.6) $\times$ 10$^{15}$  &  \nodata                                            &  (2.5 $\pm$ 1.0) $\times$ 10$^{-8}$     &  \nodata        & $N$ derived from $^{34}$S isotopologue  \\
C$^{34}$S                                 &  \nodata            &  (9.1 $\pm$ 0.5) $\times$ 10$^{13}$  &  \nodata                                            &  (1.1 $\pm$ 0.3) $\times$ 10$^{-9}$     &  0.65/730       & $T_{\mathrm{rot}}$ = $T_{\mathrm{rot}}$(C$^{33}$S) assumed   \\
C$^{33}$S                                 &  32$^{+2}_{-1}$     &  (3.4 $\pm$ 0.2) $\times$ 10$^{13}$  &  \nodata                                            &  (4.2 $\pm$ 1.1) $\times$ 10$^{-10}$    &  0.42/470       &  \\
H$_2$CS                                   &  43$^{+1}_{-1}$     &  (2.8 $\pm$ 0.1) $\times$ 10$^{14}$  & (2.8 $\pm$ 0.1) $\times$ 10$^{14}$\tablenotemark{a} &  (3.5 $\pm$ 0.9) $\times$ 10$^{-9}$     & 0.63/710        &  \\
H$_2$$^{13}$CS                            &  \nodata            &  $<$3 $\times$ 10$^{12}$             &  \nodata                                            &  $<$4 $\times$ 10$^{-11}$               & \nodata         & $T_{\mathrm{rot}}$ = $T_{\mathrm{rot}}$(H$_2$CS) assumed  \\
OCS                                       &  89$^{+1}_{-1}$     &  (1.9 $\pm$ 0.6) $\times$ 10$^{15}$  &  (1.9 $\pm$ 0.1) $\times$ 10$^{15}$                 &  (2.4 $\pm$ 0.9) $\times$ 10$^{-8}$     &  0.50/560       &  \\
O$^{13}$CS                                &  \nodata            &  (3.4 $\pm$ 1.3) $\times$ 10$^{13}$  &  \nodata                                            &  (4.3 $\pm$ 2.0) $\times$ 10$^{-10}$    &  \nodata        & $T_{\mathrm{rot}}$ = $T_{\mathrm{rot}}$(OCS) assumed \\
SO                                        &  22$^{+1}_{-1}$     &  3.4 $\times$ 10$^{15}$              &  (7.5 $\pm$ 0.4) $\times$ 10$^{15}$                 &  \nodata                                &  0.66/740       & Likely optically thick \\
SO ($^{34}$S)                             &  \nodata            &  (6.3 $\pm$ 2.9) $\times$ 10$^{15}$  &  \nodata                                            &  (7.9 $\pm$ 4.1) $\times$ 10$^{-8}$     &  \nodata        & $N$ derived from $^{34}$S isotopologue  \\
$^{34}$SO                                 &  40$^{+12}_{-7}$    &  (2.9 $\pm$ 1.0) $\times$ 10$^{14}$  &  \nodata                                            &  (3.6 $\pm$ 1.5) $\times$ 10$^{-9}$     &  0.50/560       &  \\
$^{33}$SO                                 &  53$^{+5}_{-5}$     &  (9.0 $\pm$ 1.1) $\times$ 10$^{13}$  &  \nodata                                            &  (1.1 $\pm$ 0.3) $\times$ 10$^{-9}$     &  0.48/540       &  \\
S$^{18}$O                                 &  \nodata            &  (2.2 $\pm$ 0.4) $\times$ 10$^{13}$  &  \nodata                                            &  (2.8 $\pm$ 0.9) $\times$ 10$^{-10}$     & $<$0.23/$<$260 & $T_{\mathrm{rot}}$ = 30--60 K assumed from $^{34}$SO and $^{33}$SO  \\
SO$_2$                                    &  113$^{+1}_{-1}$    &  (1.0 $\pm$ 0.1) $\times$ 10$^{16}$  &  (1.1 $\pm$ 0.1) $\times$ 10$^{16}$                 &  (1.3 $\pm$ 0.3) $\times$ 10$^{-7}$     &  0.46/520       &  \\
$^{34}$SO$_2$                             &  89$^{+5}_{-5}$     &  (4.4 $\pm$ 0.3) $\times$ 10$^{14}$  &  \nodata                                            &  (5.4 $\pm$ 1.4) $\times$ 10$^{-9}$     &  0.51/570       &  \\
$^{33}$SO$_2$                             &  112$^{+33}_{-21}$  &  (7.8 $\pm$ 1.8) $\times$ 10$^{13}$  &  \nodata                                            &  (9.8 $\pm$ 3.3) $\times$ 10$^{-10}$    &  0.51/570       &  \\
HDS$\dag$                                 &  \nodata           &  (2.5 $\pm$ 1.3) $\times$ 10$^{13}$   &  \nodata                                            &  (3.1 $\pm$ 1.8) $\times$ 10$^{-10}$    & \nodata         & $T_{\mathrm{rot}}$ = $T_{\mathrm{rot}}$(SO$_2$) assumed  \\
CH$_3$SH                                  &  \nodata           &  $<$8 $\times$ 10$^{12}$              &  \nodata                                            &  $<$1 $\times$ 10$^{-10}$               & \nodata         & $T_{\mathrm{rot}}$ = $T_{\mathrm{rot}}$(SO$_2$) assumed  \\
\enddata
\tablecomments{
Uncertainties and upper limits are at the 2$\sigma$ level and do not include systematic errors due to adopted spectroscopic constants. \\
$\dag$ indicates tentative detection. 
}
\added{
\tablenotetext{a}{Assuming ortho/para ratio of 3.}
\tablenotetext{b}{Assuming E-CH$_3$OH/A-CH$_3$OH ratio of unity \citep{Wir11}.}
}
\end{deluxetable*}

\subsection{Molecular column densities and gas temperatures} \label{sec_rd}
For the molecular for which multiple transitions with different excitation energies are detected, the column densities ($N$) and rotational temperatures ($T_{\mathrm{rot}}$) are estimated based on the rotation diagram analysis. 
The details of the formulae and the results are described in Appendix~\ref{sec_app_rd}. 
The derived values are summarized in Table~\ref{tab_N}. 

For other molecules, the rotational temperatures are assumed for the column density estimation. 
For chemically related species, for example, CH$_3$CN and $^{13}$CH$_3$CN, the $T_{\mathrm{rot}}$ derived from the rotation diagram of one species is adopted for the other (see the notes in Table~\ref{tab_N}). 
For species that show, or are expected to show, relatively extended distributions, namely H$^{13}$CO$^+$, HC$^{18}$O$^+$, CCH, CN, H$^{13}$CN, HC$^{15}$N, and c-C$_3$H$_2$, we assume that they originate from a relatively low-temperature environment and vary $T_{\mathrm{rot}}$ between 20 and 40 K to estimate the possible range of column densities. 
For SiO and H$_2$CO, HDCO, and D$_2$CO, whose rotational temperatures are unknown, we vary $T_{\mathrm{rot}}$ over a wide range from 50 to 150 K to estimate the column densities. 

For molecular species whose main isotopologues are likely to be optically thick (CH$_3$OH, SO, and CS), as well as for species whose main isotopologues are not detected (HCO$^+$ and HCN), column densities are estimated using their rare isotopologues. 
From the present observations, the following isotope ratios are derived: 
$^{12}$C/$^{13}$C = 56–69 (from O$^{12}$CS/O$^{13}$CS and $^{12}$CH$_3$CN/$^{13}$CH$_3$CN), and $^{34}$S/$^{33}$S = 3–6 (from C$^{34}$S/C$^{33}$S, $^{34}$SO/$^{33}$SO, and $^{34}$SO$_2$/$^{33}$SO$_2$). 
These values are consistent with those in the local ISM \citep{Yan23}, suggesting that the isotope ratios in HC1 have not been significantly modified by the supernova event. 
We therefore adopt $^{12}$C/$^{13}$C = 66 and $^{32}$S/$^{34}$S = 22, which correspond to the typical isotope ratios at the Galactocentric distance of RX~J1713 (7.3 kpc), as estimated from the data of \citet{Yan23}. 
We assume an uncertainty of 30$\%$ in these ratios and propagate it into the uncertainties of the column density estimates.

\added{
We also performed non-LTE calculations using RADEX \citep{vdT07}, adopting the uniform-sphere escape probability formalism. 
For the input parameters, we adopt an H$_2$ gas density of 1.3 $\times$ 10$^{7}$~cm$^{-3}$, as estimated in Section~\ref{sec_h2}, and a background temperature of 2.73~K. 
The kinetic temperatures are assumed to be equal to those listed in Table~\ref{tab_N}. 
The line intensities and widths are taken from the tables in Appendix~\ref{sec_app_lineparam}
\footnote{The following lines are used for the RADEX calculations: 
H$^{13}$CO$^+$(3-2), 
HC$^{18}$O$^+$(4-3), 
c-C$_3$H$_2$(5$_{3,2}$-4$_{4,1}$), 
H$_2$CO(5$_{1,5}$-4$_{1,4}$), 
CH$_3$OH(5$_{0}$ E--4$_{0}$ E, 5$_{2}$ A$^+$--4$_{2}$ A$^+$, and 7$_{-5}$ E--6$_{-5}$ E), 
CN(N = 3-2, J = $\frac{5}{2}$-$\frac{3}{2}$, F = $\frac{5}{2}$-$\frac{5}{2}$), 
H$^{13}$CN(3-2), 
HC$^{15}$N(3-2), 
NO(J = $\frac{7}{2}$-$\frac{5}{2}$, $\Omega$ = $\frac{1}{2}$, F = $\frac{9}{2}$$^+$-$\frac{7}{2}$$^-$), 
CH$_3$CN(14$_{0}$-13$_{0}$), 
HC$_3$N(27-26), 
SiO(6-5), 
CS(5-4), 
H$_2$CS(7$_{1,6}$-6$_{1,5}$), 
OCS(20-19), 
SO($N_J$ = 6$_{6}$-5$_{5}$),
and SO$_2$(5$_{4,2}$-6$_{3,3}$).}. 
For CH$_3$OH, we used three transitions and derived the plausible range of column densities. 
The results of non-LTE calculations are summarized in Table~\ref{tab_N}. 
Overall, the non-LTE column densities are in good agreement with the LTE estimates. 
For CS and SO, which appear to be optically thick, the isotopologue-based LTE estimates are reasonably consistent with the non-LTE results. 
For SiO, the non-LTE column density is about a factor of two lower than the LTE value. 
The SiO line profiles clearly show broad wings, and contamination from outflow-related emission components may be responsible for the discrepancy. 
}

\subsection{H$_2$ column density and fractional abundances } \label{sec_h2} 
The column density of molecular hydrogen ($N_{\mathrm{H_2}}$) is estimated from the dust continuum emission under the assumption of optically thin conditions. 
The dust temperature ($T_d$) is varied between 100 and 150~K to assess the plausible range of $N_{\mathrm{H_2}}$. 
Further details and the parameters adopted in the calculation are described in Appendix~\ref{sec_app_h2}. 
The resulting H$_2$ column densities are estimated to be 
$N_{\mathrm{H_2}}$ = (8.0 $\pm$ 2.0) $\times$ 10$^{22}$~cm$^{-2}$ for HC1 and 
$N_{\mathrm{H_2}}$ $\sim$1 $\times$ 10$^{23}$~cm$^{-2}$ for HC2. 
The molecular fractional abundances calculated using these $N(\mathrm{H_2})$ values are summarized in Table~\ref{tab_N}. 
These $N(\mathrm{H_2})$ correspond to gas number densities of 
$n_{\mathrm{H_2}}$ $\sim$1 $\times$ 10$^{7}$~cm$^{-3}$ for HC1 and 
$n_{\mathrm{H_2}}$ $\sim$2 $\times$ 10$^{7}$~cm$^{-3}$ for HC2, 
assuming a source size of 600~au.

\section{Discussion} \label{sec_disc} 
\subsection{Physical properties} \label{sec_disc_phys} 
Both of the two sources observed in this study (RX1713 HC1 and HC2) show emission lines from various high-temperature molecular gas including COMs. 
High-$E_{\rm u}$ molecular emission are spatially compact ($<500$~au) and are associated with the continuum peaks. 
Both sources are associated with dense gas ($n_{\mathrm{H_2}} \gtrsim 1 \times 10^{7}$~cm$^{-3}$). 
Excitation analyses of HC1 reveal rotational temperatures exceeding 100~K for many molecular species. 
Such temperatures are sufficient to induce the sublimation of ice mantles. 
The bolometric luminosities are estimated from spectral energy distribution (SED) data covering wavelengths from 3 to 500~$\mu$m, compiled from available databases (see Appendix~\ref{sec_app_sed} for details). 
The resulting luminosities are 260~$L_{\odot}$ for HC1 and 230~$L_{\odot}$ for HC2. 
Their SEDs have a peak in mid- to far-infrared regions, which would suggest their Class I nature. 
Based on these characteristics, we interpret these two sources as hot molecular cores associated with intermediate-mass Class~I protostars. 
This represents the first detection of hot cores in a supernova-feedback region.

\begin{figure}[tp!] 
\begin{center} 
\includegraphics[width=8.0cm]{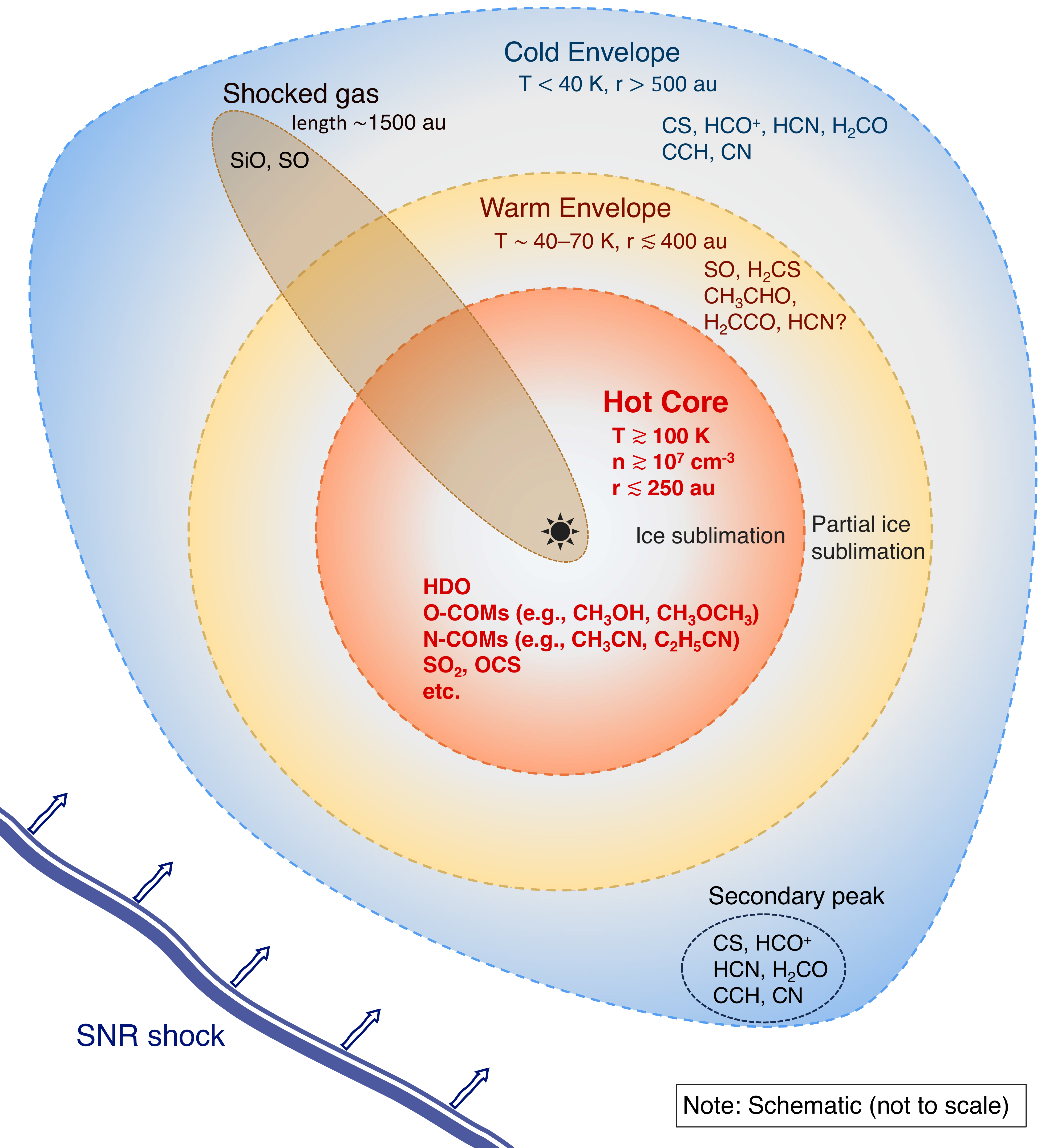} 
\caption{
Schematic illustration of the gas distribution and temperature structure in RX1713 HC1 (see Section~\ref{sec_disc_phys} for details.). 
}
\label{schematic}
\end{center}
\end{figure}

\added{Figure~\ref{schematic} illustrates the spatial distribution and temperature structure of the gas surrounding RX1713 HC1, based on the physical parameters summarized in Table~\ref{tab_N}. 
Most COMs are likely located in the hot-core region, as suggested by their high rotational temperatures and compact spatial distributions. 
SO, H$_2$CS, and some COMs (CH$_3$CHO and H$_2$CCO) exhibit somewhat lower rotational temperatures and/or slightly extended distributions, suggesting that they trace the warm envelope located just outside the hot-core region. 
In the outer cold envelope, species such as CS, which shows a low rotational temperature, as well as HCO$^+$, HCN, and H$_2$CO, which exhibit more extended spatial distributions, are likely dominant. 
CCH and CN show even more extended distributions and are also likely located in the cold region. 
On the southwestern side of HC1, a secondary peak is seen in several molecular lines (CS, HCO$^+$, HCN, H$_2$CN, CCH, and CN) as well as in the dust continuum. 
The nature of this peak is unclear at present. }

\added{The progenitor of the RX1713 SNR is located to the southeast of HC1, and strong shock waves are likely arriving from that direction. 
Dense and sharply varying dust-continuum contours are observed toward the southeastern side of HC1 (see Appendix~\ref{sec_app_img}), which may reflect compression of gas and dust caused by such SNR shocks. }

\added{From a kinematic perspective, most high-excitation molecular lines show a single velocity component and therefore likely originate in the hot-core region. 
On the other hand, several lines exhibit blue-/red-shifted wing-like features that deviate from a Gaussian profile. 
Such wing components are observed in CS, SO, H$_2$CO, and H$^{13}$CN, and are particularly prominent in SiO (see Figures~\ref{spec_B6}-\ref{spec_B7}). 
For SiO, an elongated structure extending toward the northeast from the hot-core position is clearly seen (see Appendix~\ref{sec_app_img}). 
In addition, in relatively low-excitation CH$_3$OH lines (e.g., 1$_{1}$-0$_{0}$ A$^+$, 2$_{2}$-3$_{1}$ A$^+$, 4$_{0}$-3$_{-1}$ E, 7$_{0}$-6$_{0}$ A$^+$; $E_{u}$ = 16.8, 44.7, 36.3, and 65.0 K, respectively), only red-shifted wing components are detected. 
Molecular gas exhibiting such velocity components may originate from outflows/jets or shocked gas in the vicinity of the protostar. 
Since the present observations do not include CO emission lines, a detailed discussion of the gas dynamics associated with the hot core is beyond the scope of this paper.}

\begin{figure*}[tpbh!]
\begin{center}
\includegraphics[width=18cm]{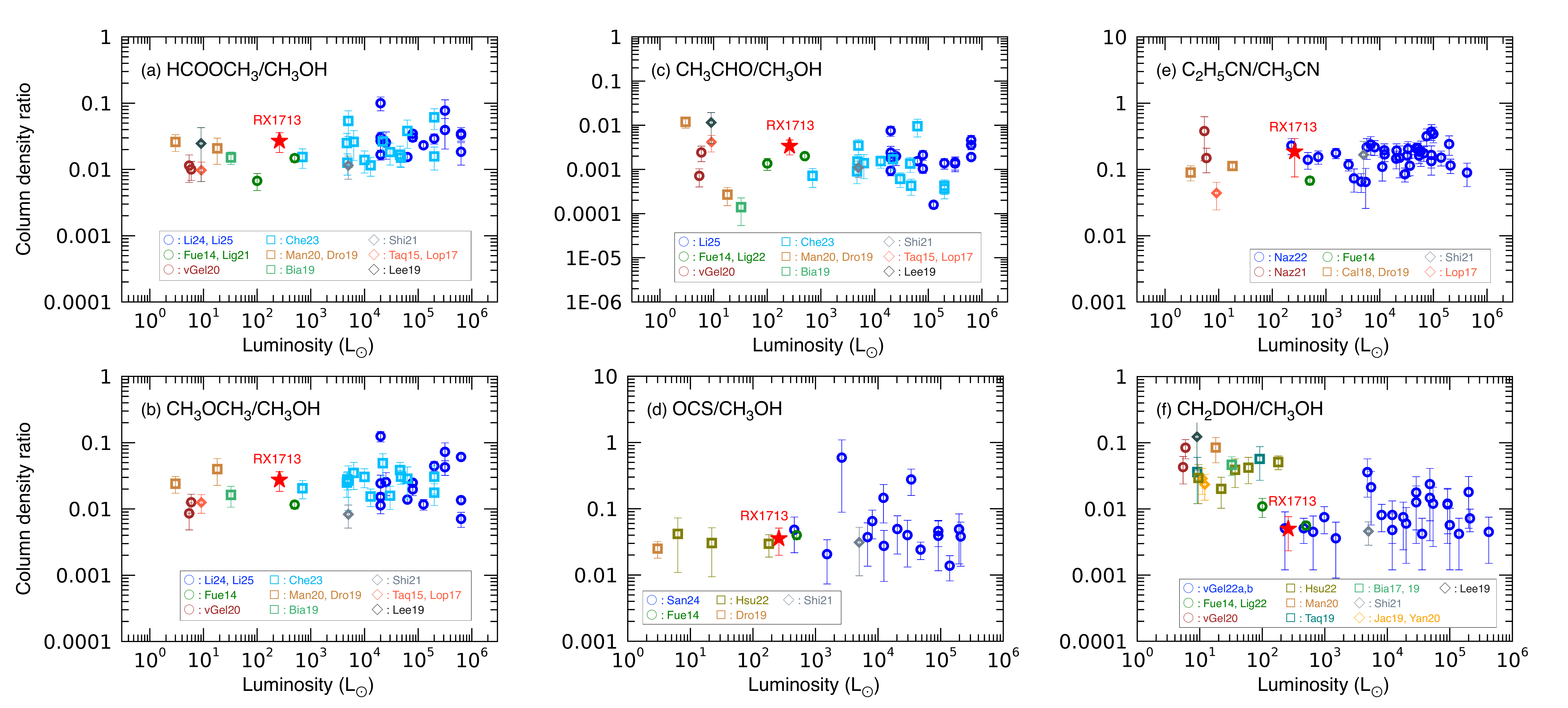}
\caption{
Comparison of the column density ratios in RX1713 HC1 with those of other hot cores and hot corinos. 
The horizontal axis represents the bolometric luminosity. 
The panels show (a) HCOOCH$_3$/CH$_3$OH, (b) CH$_3$OCH$_3$/CH$_3$OH, (c) CH$_3$CHO/CH$_3$OH, (d) OCS/CH$_3$OH, (e) C$_2$H$_5$CN/CH$_3$CN, and (f) CH$_2$DOH/CH$_3$OH. 
The red star indicates the result for RX1713 HC1 derived in this work. 
Literature values are shown with different symbols and colors as indicated in the legend. 
The corresponding references are listed below. 
High-mass hot cores: 
\citet[][Li24]{Li24}, 
\citet[][Li25]{Li25}, 
\citet[][Che23]{Che23}, 
\citet[][vGel22a]{vGel22a}, 
\citet[][vGel22b]{vGel22b}, 
\citet[][Naz22]{Naz22}, 
\citet[][San24]{San24}; 
intermediate-mass hot cores: 
\citet[][Fue14]{Fue14}, 
\citet[][Lig21]{Lig21}, 
\citet[][Lig22]{Lig22}; 
intermediate-mass hot core in the extreme outer Galaxy (WB89-789 SMM1): 
\citet[][Shi21]{ST21}; 
low-mass hot corinos: 
\citet[][Hsu22\footnote{OCS and CH$_3$OH column densities are re-estimated using their $^{13}$C isotopologue. }]{Hsu22}, 
\citet[][vGel20]{vGel20}, 
\citet[][Jac19]{Jac19}, 
\citet[][Yan20]{Yan20},  
\citet[][Taq15]{Taq15}, 
\citet[][Taq19]{Taq19}, 
\citet[][Naz21]{Naz21}, 
\citet[][Bia17]{Bia17}, 
\citet[][Bia19]{Bia19}, 
\citet[][Lop17]{Lop17}; 
low-mass hot corinos (IRAS 16293-2422 A, B): 
\citet[][Man20]{Man20}, 
\citet[][Dro19]{Dro19}, 
\citet[][Cal18]{Cal18}; 
low-mass edge-on disk (HH212): 
\citet[][Lee19]{Lee19}
. 
}
\label{abundance}
\end{center}
\end{figure*}

\subsection{Chemical properties} \label{sec_disc_chem} 
Figure~\ref{abundance} compares the chemical composition of RX1713 HC1 with those of other known hot cores and hot corinos. 
Panels (a), (b), and (c) focus on COMs, which are thought to be products of grain-surface reactions, and column density ratios normalized to CH$_3$OH are plotted. 
The normalization by CH$_3$OH is adopted because this molecule is considered to be a parental species of larger COMs \citep[e.g.,][]{Gar06}. 
As shown in the figure, the COM composition of RX1713 HC1 does not show any significant deviation from those of other hot cores/corinos.

Panel (d) of Figure~\ref{abundance} compares the OCS/CH$_3$OH ratio in order to examine the composition of sulfur-bearing molecules. 
OCS is considered to be a product of grain-surface reactions, as it has been widely detected in ice mantles through infrared observations \citep[e.g.,][]{Boo22}. 
It has also been reported that the abundances of OCS in ice mantles and in high-temperature gas are similar, suggesting that OCS in hot cores may preserve a record of solid-phase reactions involving sulfur-bearing species that occurred at earlier evolutionary stages \citep{San24}. 
Panel (e) of Figure~\ref{abundance} compares the abundance ratio of the large nitrogen-bearing molecule C$_2$H$_5$CN to its potential parental species, CH$_3$CN.
As in the case of COMs, the column density ratios of these species are comparable to those observed in other known hot cores/corinos.

Panel (f) of Figure~\ref{abundance} shows a comparison of the CH$_2$DOH/CH$_3$OH ratio in HC1 with those of other hot cores/corinos. 
Deuterium fractionation is known to proceed efficiently under cold conditions \citep[e.g.,][]{Rob03,Cec14,Taq14,Fur16}. 
This is because the key reaction triggering deuterium fractionation, $\mathrm{H_3^+ + HD \rightarrow H_2D^+ + H_2 + 232~K}$, is exothermic, and its backward reaction cannot proceed efficiently below temperatures of $\sim$20~K. 
In addition, on dust-grain surfaces, deuterium fractionation of CH$_3$OH and related species is known to proceed efficiently under cold conditions through successive H-abstraction and D-substitution reactions \citep[e.g.,][]{Hid09,Rie23}. 
As a result of these processes, the abundances of deuterated molecules observed in star-forming regions (e.g., HDCO and CH$_2$DOH) are orders of magnitude higher than the cosmic D/H ratio \citep[$\sim$2 $\times$ $10^{-5}$;][]{Lin06}. 
The CH$_2$DOH/CH$_3$OH ratio in HC1 is approximately 0.5$\%$ and does not show any significant difference from those measured in other objects with similar luminosities. 
This suggests that HC1 experienced sufficiently cold conditions during the ice-formation stage, similar to protostellar objects in typical star-forming environments.

\subsection{\added{Limited impact of supernova feedback on the molecular complexity of HC1}} \label{sec_disc_chem2} 
The above comparisons suggest that the chemical compositions of major COMs and nitrogen- and sulfur-bearing molecules in HC1, as well as the degree of CH$_3$OH deuterium fractionation, have not been significantly altered by supernova feedback. 
Why, then, has the chemical evolution in HC1 not been significantly influenced by such feedback? 
The age of RX1713 is estimated to be approximately 1600~yr, based on historical records of a guest star in Chinese chronicles \citep{Wan97} and on comparisons between X-ray observations and hydrodynamical calculations of supernova explosions \citep{Tsu16}. 
The typical ages of protostars in the hot-core or hot-corino phase are on the order of $10^4$--$10^5$~yr, and the duration of this evolutionary stage is thought to be comparable \citep[e.g.,][]{Cas12}. 
This suggests that HC1 and HC2 were already formed as protostars and had developed hot-core regions prior to the supernova explosion. 
During the prestellar stage, these sources were likely exposed to strong stellar winds from the massive star that later became the supernova progenitor \citep{San10}. 
However, the degree of CH$_3$OH deuterium fractionation in HC1 suggests that such stellar winds did not significantly affect the physical conditions during the ice-formation stage. 

A shock velocity of $\sim$4000~km s$^{-1}$ is observed near HC1 \citep{Tsu16}. 
However, such shocks may have been decelerated through interactions with the surrounding high-density medium and may not yet have reached the position of HC1, as implied by the persistence of the hot-core region (also see the discussion of shock deceleration in \citeauthor{San10} \citeyear{San10}). 

The gas column density from the outer edge to the center of HC1 is estimated to be $N(\mathrm{H_2})$ $\sim$4 $\times$ 10$^{22}$ cm$^{-2}$. 
Considering the penetration depths of primary and secondary cosmic-ray particles, cosmic rays with energies above $\sim$1 MeV can penetrate into the hot-core region \citep{Pad18}. 
Similarly, high-energy photons with energies above $\sim$10 keV can penetrate, whereas photons with energies below $\sim$1 keV are unlikely to reach the source center \citep{Bet11}. 
However, magnetic fields in molecular clouds surrounding SNRs may be amplified through interactions between the supernova shock and the ambient medium \citep[e.g.,][]{Uch07, Ino12}. 
In such cases, the penetration efficiency of cosmic rays into molecular clouds would be significantly suppressed \citep[e.g.,][]{Max12,Cel19,San20}. 
The hot-core region may therefore be partially shielded from cosmic rays by magnetic fields amplified by the supernova feedback. 

Another important factor is the duration of exposure to energetic particles and photons. 
HC1 is located at a projected distance of approximately 10~pc from 1WGAJ1713.4$-$3949, which is considered to be the progenitor of RX1713, and this location roughly corresponds to the outer edge of the observed supernova shell (see Figure~\ref{rx1713}). 
The region surrounding HC1 has therefore likely only recently begun to be exposed to intense energetic particles and photons. 

In Appendix \ref{sec_app_networkmodel}, we perform astrochemical simulations to investigate the impact of an enhanced cosmic-ray ionization rate on the destruction of COMs in hot cores under simplified physical conditions. 
Our calculations show that, for an ionization rate below 10$^{-15}$ s$^{-1}$, the gas-phase molecular abundances do not change significantly over a timescale of $\sim$1000 years. 
Given the location of HC1 and the possible suppression of cosmic-ray penetration by amplified magnetic fields, the elapsed time since the onset of exposure to enhanced high-energy particles and photons may be shorter than $\sim$1000 years.
A longer period of exposure to such a harsh environment would be required for the COM composition of HC1 to change significantly.

\section{Summary} \label{sec_sum} 
We report the detection of two hot molecular cores located inside the shell of the young supernova remnant RX~J1713.7$-$3946 based on ALMA observations. 
Our observations reveal that the chemical composition of HC1 is indistinguishable from those of hot cores and hot corinos in more ordinary environments. 
HC1 is located near the outer edge of the supernova shell, and the surrounding region has likely begun to be exposed to such a harsh environment only recently. 
The elapsed time since the onset of exposure to high-energy particles and photons may be too short for the chemical composition of the hot core to be significantly altered, and/or the hot-core region may be shielded by magnetic fields amplified by supernova feedback, which could suppress the penetration of enhanced cosmic rays. 
%
We note that the present results represent a case study of HC1 and do not necessarily imply that supernova feedback in RX1713 has no impact on the chemical evolution of other star-forming cores in the same region. 
The other newly discovered hot core, HC2, is located at a projected distance of approximately 4~pc from the progenitor, a factor of 2.5 closer than HC1. 
HC2 may therefore have been exposed to high-energy particles and photons for a longer period. 
A detailed analysis of HC2 will be presented in a forthcoming paper.

\begin{acknowledgments}
This paper makes use of the following ALMA data: ADS/JAO.ALMA$\#$2024.1.00402.S. 
ALMA is a partnership of ESO (representing its member states), NSF (USA) and NINS (Japan), together with NRC (Canada), NSTC and ASIAA (Taiwan), and KASI (Republic of Korea), in cooperation with the Republic of Chile. 
The Joint ALMA Observatory is operated by ESO, AUI/NRAO and NAOJ.
This work has made extensive use of the CDMS and JPL data. 
This research has made use of the NASA/IPAC Infrared Science Archive, which is funded by the National Aeronautics and Space Administration and operated by the California Institute of Technology. 
\added{This work was supported by JSPS KAKENHI grant Nos. 20H05845, 24H00246, 25K07364, 25K07367, 26KF0052, 26K00761, and 26K22372. }
T. S. acknowledges support from the Uchida Energy Science Promotion Foundation. 
Y.O. acknowledges NAOJ for the ALMA Joint Scientific Research Program (2024-27B). 
\added{Finally, we would like to thank an anonymous referee for insightful comments, which substantially improved this paper. }
\end{acknowledgments}

\software{CASA \citep{McM07})}
\facility{ALMA, IRSA, Herschel, WISE, XMM}


\appendix

\restartappendixnumbering
\section{Measured line parameters} \label{sec_app_lineparam}
Tables \ref{tab_lines1}--\ref{tab_lines7} summarize measured line parameters (see Section \ref{sec_spc} for details). 
We have measured the peak brightness temperature ($T_{b}$), the line FWHMs (full widths at half maximums, $\Delta$$V$), the integrated intensities ($\int T_{b} dV$), and the LSR velocity ($V_{LSR}$) for each line by fitting a Gaussian profile. 
For spectral lines for which a Gaussian does not fit well, their integrated intensities are calculated by directly integrating the spectrum over the frequency region of emission. 
The tables also list the estimated upper limits on important non-detection lines. 
The quoted uncertainties and upper limits correspond to the 2$\sigma$ level. 
Upper limits are estimated assuming $\Delta$$V$ = 4 km s$^{-1}$. 
The signal-to-noise ratios (S/N) of the peak brightness temperatures are also provided.

\startlongtable
\begin{deluxetable*}{ l c c c c c c c c c }
\tablecaption{Line Parameters for H$^{13}$CO$^+$, HC$^{18}$O$^+$, CCH, c-C$_3$H$_2$, H$_2$CO, HDCO, D$_2$CO, and HDO \label{tab_lines1}}
\tablewidth{0pt}
\tabletypesize{\scriptsize} 
\tablehead{
\colhead{Molecule}                       & \colhead{Transition}                                                                    &       \colhead{$E_u$} &       \colhead{Frequency} &        \colhead{$T_{b}$} &     \colhead{$\Delta$$V$} &     \colhead{$\int T_{b} dV$} &       \colhead{$V_{LSR}$} &        \colhead{S/N} &       \colhead{Note} \\
\colhead{ }                                  & \colhead{ }                                                                                     &        \colhead{(K)}    &         \colhead{(GHz)}    &             \colhead{(K)}     &          \colhead{(km/s)}       &             \colhead{(K km/s)}     &          \colhead{(km/s)} &           \colhead{} &           \colhead{}
}
\startdata
 H$^{13}$CO$^+$                          &  3--2                                                                                                                    &   25       & 260.25534       &   5.09 $\pm$   0.06  &    3.5          & 19.23 $\pm$ 0.24     &  -8.3           &       96.0 &    \nodata \\
 HC$^{18}$O$^+$                          &  4--3                                                                                                                    &   41       & 340.63069       &   0.43 $\pm$   0.06  &    3.5          & 1.63 $\pm$ 0.19      &  -8.5           &        7.9 &    \nodata \\
 CCH                                     &  N = 4--3, J = $\frac{9}{2}$--$\frac{7}{2}$, F = 5--4                                                                    &   42       & 349.33771       &   3.20 $\pm$   0.06  &    3.9          & 13.26 $\pm$ 0.26     &  -8.9           &       52.5 &        (1) \\
 CCH                                     &  N = 4--3, J = $\frac{7}{2}$--$\frac{5}{2}$, F = 4--3                                                                    &   42       & 349.39928       &   2.67 $\pm$   0.05  &    4.7          & 13.46 $\pm$ 0.27     &  -8.5           &       43.9 &        (1) \\
 c-C$_3$H$_2$                            &  5$_{3,2}$--4$_{4,1}$                                                                                                    &   45       & 260.47975       &   0.14 $\pm$   0.08  &    1.9          & 0.28 $\pm$ 0.17      &  -7.7           &        2.7 &        (2) \\
 H$_2$CO                                 &  5$_{1,5}$--4$_{1,4}$                                                                                                    &   62       & 351.76864       &  16.48 $\pm$   0.05  &    4.6          & 79.89 $\pm$ 0.23     &  -8.4           &      274.6 &    \nodata \\
\enddata
\tablecomments{
(1) Blend of multiple components with similar spectroscopic constants. 
(2) Tentative detection. 
}
\end{deluxetable*}

\startlongtable
\begin{deluxetable*}{ l c c c c c c c c c }
\tablecaption{Line Parameters for N-bearing molecules \label{tab_lines2}}
\tablewidth{0pt}
\tabletypesize{\scriptsize} 
\tablehead{
\colhead{Molecule}                       & \colhead{Transition}                                                                    &       \colhead{$Eu$} &       \colhead{Frequency} &        \colhead{$T_{b}$} &     \colhead{$\Delta$$V$} &     \colhead{$\int T_{b} dV$} &       \colhead{$V_{LSR}$} &        \colhead{S/N} &       \colhead{Note} \\
\colhead{ }                                  & \colhead{ }                                                                                     &        \colhead{(K)}    &         \colhead{(GHz)}    &             \colhead{(K)}     &          \colhead{(km/s)}       &             \colhead{(K km/s)}     &          \colhead{(km/s)} &           \colhead{} &           \colhead{}
}
\startdata
 H$^{13}$CN                              &  3--2                                                                                                                    &   25       & 259.01180       &   9.32 $\pm$   0.04  &    6.3          & 62.13 $\pm$ 0.34     &  -8.7           &      175.9 &    \nodata \\
 HC$^{15}$N                              &  3--2                                                                                                                    &   25       & 258.15700       &   4.96 $\pm$   0.05  &    5.7          & 30.07 $\pm$ 0.33     &  -8.7           &       95.4 &    \nodata \\
  HCCN                                    &  11$_{12}$--10$_{11}$                                                                                                    &   70       & 241.64386       &             $<$0.10  &    \nodata      &             $<$0.2   &    \nodata      &    \nodata &    \nodata \\
 HC$_3$N                                 &  27--26                                                                                                                  &  165       & 245.60632       &   3.04 $\pm$   0.04  &    6.1          & 19.86 $\pm$ 0.26     &  -9.0           &       63.4 &    \nodata \\
 CN                                      &  N = 3--2, J = $\frac{5}{2}$--$\frac{5}{2}$, F = $\frac{7}{2}$--$\frac{7}{2}$                                            &   33       & 339.51664       &   0.25 $\pm$   0.06  &    3.0          & 0.82 $\pm$ 0.17      &  -8.4           &        4.6 &    \nodata \\
 CN                                      &  N = 3--2, J = $\frac{5}{2}$--$\frac{3}{2}$, F = $\frac{5}{2}$--$\frac{5}{2}$                                            &   33       & 340.00813       &   0.54 $\pm$   0.07  &    2.5          & 1.46 $\pm$ 0.17      &  -8.7           &        9.9 &    \nodata \\
\enddata
\tablecomments{
(1) Blend of multiple components with similar spectroscopic constants. 
(2) Tentative detection. 
}
\end{deluxetable*}

\startlongtable
\begin{deluxetable*}{ l c c c c c c c c c }
\tablecaption{Line Parameters for Si- and S-bearing molecules \label{tab_lines3}}
\tablewidth{0pt}
\tabletypesize{\scriptsize} 
\tablehead{
\colhead{Molecule}                       & \colhead{Transition}                                                                    &       \colhead{$Eu$} &       \colhead{Frequency} &        \colhead{$T_{b}$} &     \colhead{$\Delta$$V$} &     \colhead{$\int T_{b} dV$} &       \colhead{$V_{LSR}$} &        \colhead{S/N} &       \colhead{Note} \\
\colhead{ }                                  & \colhead{ }                                                                                     &        \colhead{(K)}    &         \colhead{(GHz)}    &             \colhead{(K)}     &          \colhead{(km/s)}       &             \colhead{(K km/s)}     &          \colhead{(km/s)} &           \colhead{} &           \colhead{}
}
\startdata
 SiO                                     &  6--5                                                                                                                    &   44       & 260.51801       &   3.71 $\pm$   0.11  &    6.9          & 39.09 $\pm$ 0.54     &  -8.7           &       70.0 &    \nodata \\
 HDS                                     &  2$_{1,1}$--2$_{0,2}$                                                                                                    &   47       & 257.78141       &   0.14 $\pm$   0.08  &    2.2          & 0.31 $\pm$ 0.17      &  -8.5           &        2.6 &        (2) \\
 CS                                      &  5--4                                                                                                                    &   35       & 244.93556       &  24.70 $\pm$   0.10  &    4.9          & 145.25 $\pm$ 0.27    &  -8.1           &      514.8 &    \nodata \\
 C$^{33}$S                               &  5--4                                                                                                                    &   35       & 242.91361       &   2.14 $\pm$   0.10  &    4.9          & 11.53 $\pm$ 0.29     &  -7.7           &       41.3 &    \nodata \\
 C$^{33}$S                               &  7--6                                                                                                                    &   65       & 340.05257       &   1.53 $\pm$   0.11  &    5.3          & 8.81 $\pm$ 0.29      &  -8.0           &       27.9 &    \nodata \\
 C$^{34}$S                               &  7--6                                                                                                                    &   65       & 337.39646       &   4.24 $\pm$   0.12  &    5.2          & 23.94 $\pm$ 0.31     &  -8.1           &       70.7 &    \nodata \\
\enddata
\tablecomments{
(1) Blend of multiple components with similar spectroscopic constants. 
(2) Tentative detection. 
}
\end{deluxetable*}

\startlongtable
\begin{deluxetable*}{ l c c c c c c c c c }
\tablecaption{Line Parameters for CH$_3$OH, $^{13}$CH$_3$OH, and CH$_2$DOH \label{tab_lines4}}
\tablewidth{0pt}
\tabletypesize{\scriptsize} 
\tablehead{
\colhead{Molecule}                       & \colhead{Transition}                                                                    &       \colhead{$Eu$} &       \colhead{Frequency} &        \colhead{$T_{b}$} &     \colhead{$\Delta$$V$} &     \colhead{$\int T_{b} dV$} &       \colhead{$V_{LSR}$} &        \colhead{S/N} &       \colhead{Note} \\
\colhead{ }                                  & \colhead{ }                                                                                     &        \colhead{(K)}    &         \colhead{(GHz)}    &             \colhead{(K)}     &          \colhead{(km/s)}       &             \colhead{(K km/s)}     &          \colhead{(km/s)} &           \colhead{} &           \colhead{}
}
\startdata
 CH$_3$OH                                &  25$_{3}$ A$^-$--25$_{2}$ A$^+$                                                                                          &  804       & 241.58876       &   0.27 $\pm$   0.05  &    5.5          & 1.62 $\pm$ 0.26      &  -7.8           &        5.3 &    \nodata \\
 CH$_3$OH                                &  5$_{0}$ E--4$_{0}$ E                                                                                                    &   48       & 241.70016       &   8.70 $\pm$   0.06  &    3.9          & 36.31 $\pm$ 0.25     &  -8.2           &      167.2 &    \nodata \\
 CH$_3$OH                                &  5$_{-1}$ E--4$_{-1}$ E                                                                                                  &   40       & 241.76723       &   9.19 $\pm$   0.09  &    4.4          & 42.72 $\pm$ 0.39     &  -8.3           &      108.1 &    \nodata \\
 CH$_3$OH                                &  5$_{0}$ A$^+$--4$_{0}$ A$^+$                                                                                            &   35       & 241.79135       &   9.36 $\pm$   0.05  &    4.7          & 47.11 $\pm$ 0.26     &  -8.4           &      180.0 &    \nodata \\
 CH$_3$OH                                &  5$_{3}$ A$^+$--4$_{3}$ A$^+$                                                                                            &   85       & 241.83272       &   7.84 $\pm$   0.05  &    5.4          & 44.72 $\pm$ 0.25     &  -8.2           &      150.8 &        (1) \\
 CH$_3$OH                                &  5$_{2}$ A$^-$--4$_{2}$ A$^-$                                                                                            &   73       & 241.84228       &   8.15 $\pm$   0.05  &    5.4          & 46.92 $\pm$ 0.29     &  -9.1           &      156.8 &        (1) \\
\enddata
\tablecomments{
(1) Blend of multiple components with similar spectroscopic constants. 
(2) Tentative detection. 
}
\end{deluxetable*}

\startlongtable
\begin{deluxetable*}{ l c c c c c c c c c }
\tablecaption{Line Parameters for HCOOCH$_3$ \label{tab_lines5}}
\tablewidth{0pt}
\tabletypesize{\scriptsize} 
\tablehead{
\colhead{Molecule}                       & \colhead{Transition}                                                                    &       \colhead{$Eu$} &       \colhead{Frequency} &        \colhead{$T_{b}$} &     \colhead{$\Delta$$V$} &     \colhead{$\int T_{b} dV$} &       \colhead{$V_{LSR}$} &        \colhead{S/N} &       \colhead{Note} \\
\colhead{ }                                  & \colhead{ }                                                                                     &        \colhead{(K)}    &         \colhead{(GHz)}    &             \colhead{(K)}     &          \colhead{(km/s)}       &             \colhead{(K km/s)}     &          \colhead{(km/s)} &           \colhead{} &           \colhead{}
}
\startdata
 HCOOCH$_3$                              &  19$_{5,14}$--18$_{5,13}$ E                                                                                              &  130       & 242.87157       &   0.52 $\pm$   0.06  &    3.9          & 2.17 $\pm$ 0.21      &  -7.7           &       10.1 &    \nodata \\
 HCOOCH$_3$                              &  19$_{5,14}$--18$_{5,13}$ A                                                                                              &  130       & 242.89603       &   0.51 $\pm$   0.06  &    4.4          & 2.40 $\pm$ 0.24      &  -7.6           &        9.8 &    \nodata \\
 HCOOCH$_3$                              &  19$_{4,15}$--18$_{4,14}$ A $\nu_t$ = 1                                                                                  &  313       & 244.06667       &   0.13 $\pm$   0.06  &    3.0          & 0.41 $\pm$ 0.18      &  -7.3           &        2.6 &        (2) \\
 HCOOCH$_3$                              &  20$_{12,8}$--19$_{12,7}$ A $\nu_t$ = 1                                                                                  &  407       & 244.19830       &   0.14 $\pm$   0.07  &    2.0          & 0.29 $\pm$ 0.16      &  -7.3           &        2.9 &        (2) \\
 HCOOCH$_3$                              &  20$_{4,17}$--19$_{4,16}$ E                                                                                              &  135       & 244.58034       &   0.56 $\pm$   0.05  &    4.1          & 2.41 $\pm$ 0.20      &  -8.0           &       11.6 &    \nodata \\
 HCOOCH$_3$                              &  20$_{4,17}$--19$_{4,16}$ A                                                                                              &  135       & 244.59405       &   0.54 $\pm$   0.06  &    3.5          & 2.03 $\pm$ 0.18      &  -7.5           &       11.2 &    \nodata \\
\enddata
\tablecomments{
(1) Blend of multiple components with similar spectroscopic constants. 
(2) Tentative detection. 
}
\end{deluxetable*}

\startlongtable
\begin{deluxetable*}{ l c c c c c c c c c }
\tablecaption{Line Parameters for CH$_3$OCH$_3$ and C$_2$H$_5$OH \label{tab_lines6}}
\tablewidth{0pt}
\tabletypesize{\scriptsize} 
\tablehead{
\colhead{Molecule}                       & \colhead{Transition}                                                                    &       \colhead{$Eu$} &       \colhead{Frequency} &        \colhead{$T_{b}$} &     \colhead{$\Delta$$V$} &     \colhead{$\int T_{b} dV$} &       \colhead{$V_{LSR}$} &        \colhead{S/N} &       \colhead{Note} \\
\colhead{ }                                  & \colhead{ }                                                                                     &        \colhead{(K)}    &         \colhead{(GHz)}    &             \colhead{(K)}     &          \colhead{(km/s)}       &             \colhead{(K km/s)}     &          \colhead{(km/s)} &           \colhead{} &           \colhead{}
}
\startdata
 CH$_3$OCH$_3$                           &  5$_{3,2}$--4$_{2,3}$ EE                                                                                                 &   26       & 241.52872       &   0.58 $\pm$   0.05  &    5.5          & 3.37 $\pm$ 0.26      &  -8.8           &       11.1 &        (1) \\
 CH$_3$OCH$_3$                           &  13$_{1,13}$--12$_{0,12}$ EE                                                                                             &   81       & 241.94654       &   1.22 $\pm$   0.06  &    3.4          & 4.43 $\pm$ 0.20      &  -7.8           &       23.4 &        (1) \\
 CH$_3$OCH$_3$                           &  14$_{1,14}$--13$_{0,13}$ EE                                                                                             &   93       & 258.54906       &   1.72 $\pm$   0.07  &    2.8          & 5.05 $\pm$ 0.19      &  -7.7           &       33.1 &        (1) \\
 CH$_3$OCH$_3$                           &  17$_{5,12}$--17$_{4,13}$ EE                                                                                             &  175       & 259.31195       &   0.28 $\pm$   0.04  &    8.3          & 2.42 $\pm$ 0.32      &  -8.5           &        5.2 &        (1) \\
 CH$_3$OCH$_3$                           &  6$_{3,4}$--5$_{2,3}$ EE                                                                                                 &   32       & 259.48973       &   0.50 $\pm$   0.07  &    3.2          & 1.71 $\pm$ 0.19      &  -7.5           &        9.4 &        (1) \\
 CH$_3$OCH$_3$                           &  6$_{3,4}$--5$_{2,3}$ AA                                                                                                 &   32       & 259.49375       &   0.33 $\pm$   0.09  &    2.0          & 0.71 $\pm$ 0.16      &  -7.2           &        5.7 &    \nodata \\
\enddata
\tablecomments{
(1) Blend of multiple components with similar spectroscopic constants. 
(2) Tentative detection. 
}
\end{deluxetable*}

\startlongtable
\begin{deluxetable*}{ l c c c c c c c c c }
\tablecaption{Line Parameters for HCOOH, CH$_3$CHO, H$_2$CCO, and c-C$_2$H$_4$O \label{tab_lines7}}
\tablewidth{0pt}
\tabletypesize{\scriptsize} 
\tablehead{
\colhead{Molecule}                       & \colhead{Transition}                                                                    &       \colhead{$Eu$} &       \colhead{Frequency} &        \colhead{$T_{b}$} &     \colhead{$\Delta$$V$} &     \colhead{$\int T_{b} dV$} &       \colhead{$V_{LSR}$} &        \colhead{S/N} &       \colhead{Note} \\
\colhead{ }                                  & \colhead{ }                                                                                     &        \colhead{(K)}    &         \colhead{(GHz)}    &             \colhead{(K)}     &          \colhead{(km/s)}       &             \colhead{(K km/s)}     &          \colhead{(km/s)} &           \colhead{} &           \colhead{}
}
\startdata
 \textit{trans}-HCOOH                    &  12$_{1,12}$--11$_{1,11}$                                                                                                &   84       & 257.97501       &   0.30 $\pm$   0.06  &    3.1          & 0.97 $\pm$ 0.19      &  -7.3           &        5.7 &    \nodata \\
 \textit{trans}-HCOOH                    &  15$_{8,7}$--14$_{8,6}$                                                                                                  &  333       & 337.44421       &   0.15 $\pm$   0.06  &    3.7          & 0.60 $\pm$ 0.23      &  -9.0           &        4.4 &        (1) \\
 \textit{trans}-HCOOH                    &  15$_{4,12}$--14$_{4,11}$                                                                                                &  181       & 338.14384       &   0.23 $\pm$   0.08  &    2.1          & 0.50 $\pm$ 0.16      &  -6.5           &        3.8 &    \nodata \\
 \textit{trans}-HCOOH                    &  15$_{3,13}$--14$_{3,12}$                                                                                                &  158       & 338.20186       &   0.21 $\pm$   0.06  &    4.9          & 1.07 $\pm$ 0.23      &  -7.1           &        4.9 &    \nodata \\
 \textit{trans}-HCOOH                    &  15$_{4,11}$--14$_{4,10}$                                                                                                &  181       & 338.24882       &   0.19 $\pm$   0.06  &    3.4          & 0.70 $\pm$ 0.20      &  -7.1           &        4.6 &    \nodata \\
 \textit{cis}-HCOOH                      &  11$_{3,8}$--10$_{3,7}$                                                                                                  &  102       & 242.36590       &   0.15 $\pm$   0.10  &    1.3          & 0.21 $\pm$ 0.15      &  -8.7           &        2.9 &        (2) \\
\enddata
\tablecomments{
(1) Blend of multiple components with similar spectroscopic constants. 
(2) Tentative detection. 
}
\end{deluxetable*}

\restartappendixnumbering
\section{Distribution of molecular line emission} \label{sec_app_img}
\begin{figure*}[tp!]
\begin{center}
\includegraphics[width=17.5cm]{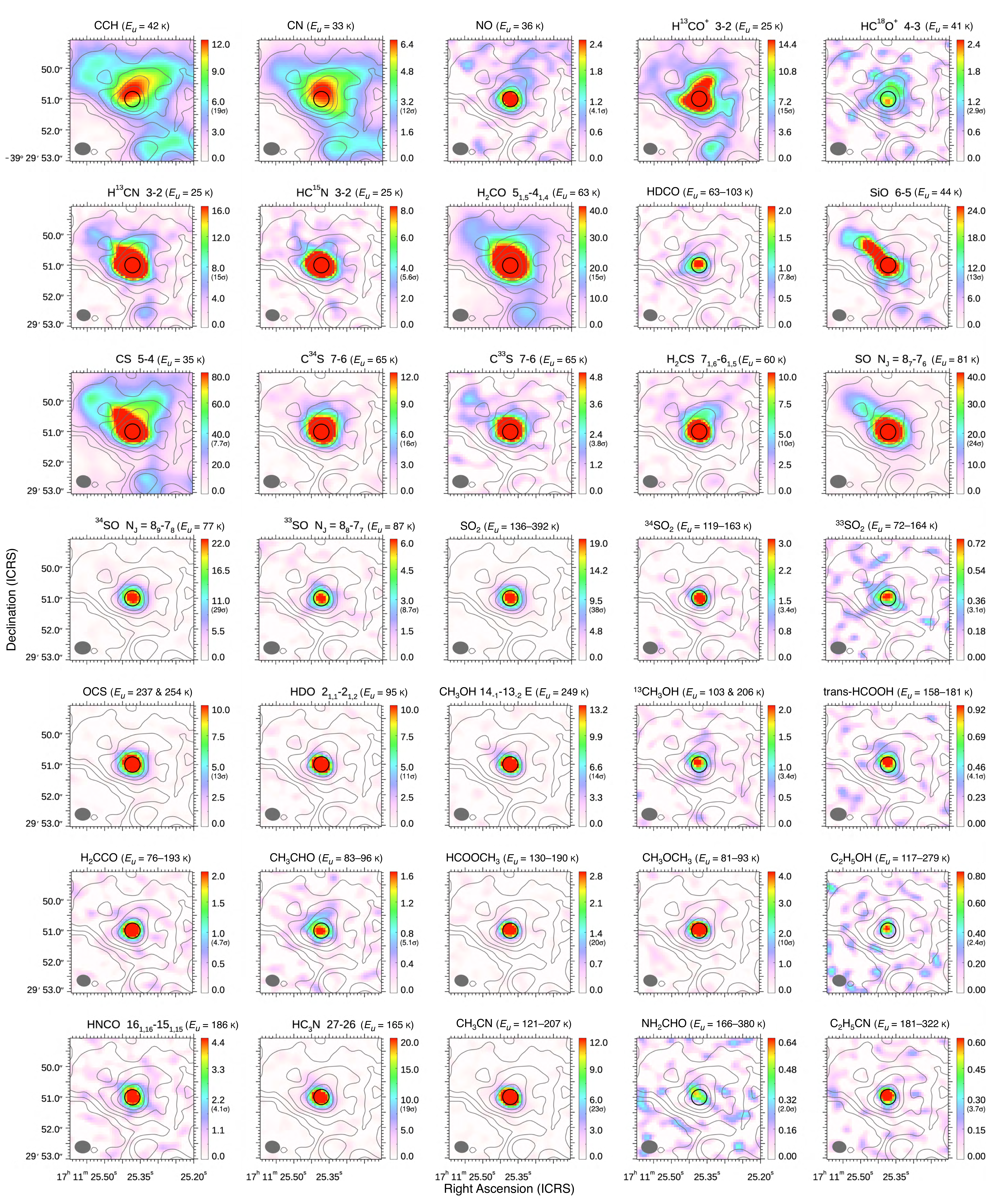}
\caption{
Integrated intensity distributions of molecular line emission. 
\added{The noise level corresponding to the midpoint of the color bar (cyan) is indicated in parentheses. }
Contours represent the 1.2~mm continuum and the contour levels are 5$\sigma$, 10$\sigma$, 20$\sigma$, 40$\sigma$, 100$\sigma$, 200$\sigma$ of the rms noise (0.02 mJy/beam). 
The spectra discussed in the text are extracted from the region indicated by the black open circle. 
The synthesized beam size is shown by the gray filled circle in each panel. 
}
\label{images}
\end{center}
\end{figure*}

Figure~\ref{images} shows synthesized images of the continuum and molecular line emission. 
The images are constructed by integrating the spectral data over the velocity ranges where the emission is detected. 
For some molecular species, multiple emission lines are stacked to improve the signal-to-noise ratio. 

\restartappendixnumbering
\section{Derivation of column densities and temperatures} \label{sec_app_calc}
\subsection{Rotation diagram analysis} \label{sec_app_rd}
\begin{figure*}[tp!]
\begin{center}
\includegraphics[width=18cm]{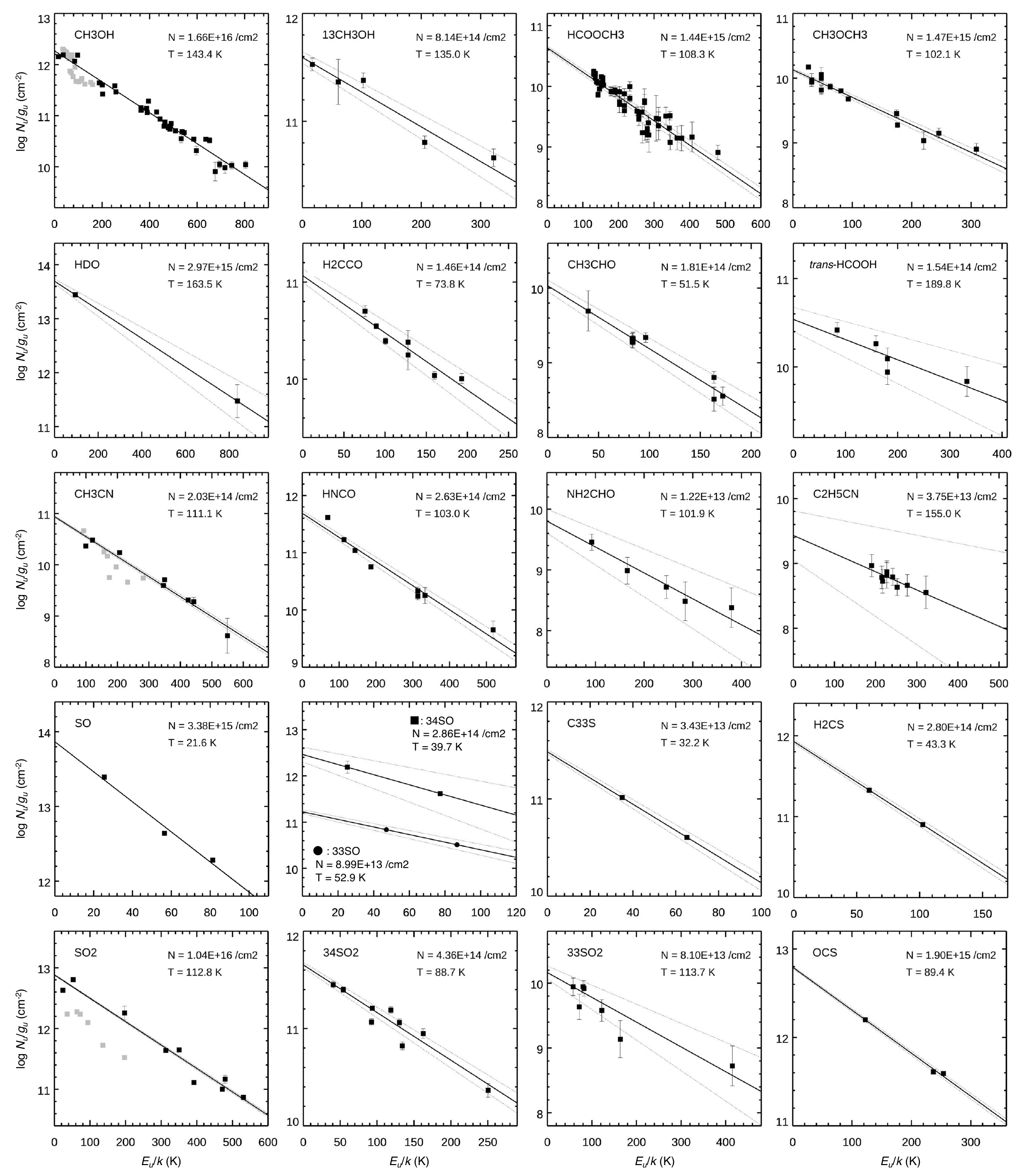}
\caption{
Results of rotation diagram analyses. 
Upper limit points are indicated by the downward arrows. 
The solid lines represent the fitted straight line, while the dashed lines indicate the acceptable fits within 2$\sigma$ level. 
Derived column densities and rotational temperatures are shown in each panel. 
The gray squares in the CH$_3$OH, CH$_3$CN, and SO$_2$ panels are excluded from the fit because of their large $S\mu^{2}$ values. 
See Section \ref{sec_rd} for details. 
}
\label{rd}
\end{center}
\end{figure*}

Figure~\ref{rd} shows the results of the excitation analysis of the molecular line emission. 
We use the following formulae based on the standard treatment of the rotation diagram method \citep[e.g., ][]{Sut95}: 
\begin{equation}
\log \left(\frac{ N_{u} }{ g_{u} } \right) = - \left(\frac {\log e}{T_{\mathrm{rot}}} \right) \left(\frac{E_{u}}{k} \right) + \log \left(\frac{N}{Q(T_{\mathrm{rot}})} \right),  \label{Eq_rd1}
\end{equation}
where 
\begin{equation}
\frac{ N_{u} }{ g_{u} } = \frac{ 3 k \int T_{\mathrm{b}} dV }{ 8 \pi^{3} \nu S \mu^{2} }, \label{Eq_rd2} \\ 
\end{equation}
and $N_{u}$ is a column density of molecules in the upper energy level, $g_{u}$ is the degeneracy of the upper level, $k$ is the Boltzmann constant, $\int T_{\mathrm{b}} dV$ is the integrated intensity estimated from the observations, $\nu$ is the transition frequency, $S$ is the line strength, $\mu$ is the dipole moment, $T_{\mathrm{rot}}$ is the rotational temperature, $E_{u}$ is the upper state energy, $N$ is the total column density, and $Q(T_{\mathrm{rot}})$ is the partition function at $T_{\mathrm{rot}}$. 
We assume an optically thin condition and the local thermodynamic equilibrium (LTE). 
All the spectroscopic parameters required in the analysis are extracted from the CDMS or JPL databases.

\subsection{Derivation of the H$_2$ column density from the dust continuum} \label{sec_app_h2} 
We use the following equation to calculate $N_{\mathrm{H_2}}$ based on the standard treatment of optically thin dust emission: 
\begin{equation}
N_{\mathrm{H_2}} = \frac{F_{\nu} / \Omega}{2 \kappa_{\nu} B_{\nu}(T_{d}) Z \overline{M}_w m_{\mathrm{H}}} \label{Eq_h2}, 
\end{equation}
where 
$F_{\nu}/\Omega$ is the continuum flux density per beam solid angle as estimated from the observations, 
$\kappa_{\nu}$ is the mass absorption coefficient of bare dust grains at 1.2 and 0.87~mm as taken from \citet{Oss94} and we here use 2.27 cm$^2$ g$^{-1}$ for 1.2~mm and 3.53 cm$^2$ g$^{-1}$ for 0.87~mm, 
$T_{d}$ is the dust temperature and $B_{\nu}(T_{d})$ is the Planck function, 
$Z$ is the dust-to-gas mass ratio and we use a canonical value of 0.008, 
$\overline{M}_w$ is the mean atomic mass per hydrogen \citep[1.41, according to][]{Cox00}, 
and $m_{\mathrm{H}}$ is the hydrogen mass. 

The dust temperature ($T_d$) is varied between 100 and 150 K to estimate the plausible range of $N_{\mathrm{H_2}}$. 
This temperature range corresponds to the mean value and standard deviation of the rotational temperatures derived for HDO, CH$_3$OH, HCOOCH$_3$, CH$_3$OCH$_3$, CH$_3$CN, NH$_2$CHO, C$_2$H$_5$CN, and SO$_2$, which are likely to originate from the hot-core region. 
The continuum brightness of HC1 is measured to be 13.7 mJy beam$^{-1}$ for 1.2~mm and 28.2 mJy beam$^{-1}$ for 0.87~mm (continuum data convolved to a 0$\farcs$54 beam size is used). 
Based on this continuum brightness, we obtain 
$N_{\mathrm{H_2}}$ = (9.2 $\pm$ 1.9) $\times$ 10$^{22}$ cm$^{-2}$ for 1.2~mm and 
$N_{\mathrm{H_2}}$ = (6.7 $\pm$ 1.5) $\times$ 10$^{22}$ cm$^{-2}$ for 0.87~mm. 
In this study, we adopt the average of these values and use $N_{\mathrm{H_2}} = (8.0 \pm 2.0) \times 10^{22}$ cm$^{-2}$ for the calculation of molecular abundances. 
It should be noted that an important source of uncertainty arises from the dust opacity. 
Taking into account the sublimation of ice mantles within the hot-core region, we adopt the dust opacity for bare dust grains estimated by \citet{Oss94}. 
If instead the opacity of thin ice-coated grains is adopted, the derived $N_{\mathrm{H_2}}$ would be approximately a factor of two larger. 
Using the above $N_{\mathrm{H_2}}$ together with \added{the size of the region used to estimate the continuum flux (about 600~au)}, the gas number density is estimated to be $n_{\mathrm{H_2}}$ = $\sim$1 $\times$ 10$^{7}$ cm$^{-3}$. 
The total mass of hot gas contained in HC1 is thus estimated to be about 0.01 M$_{\odot}$. 

For HC2, the continuum brightness is measured to be 9.5 mJy beam$^{-1}$ at 1.2~mm, based on continuum data with a beam size of $0\farcs39 \times 0\farcs47$. 
Assuming the same conditions as those adopted for HC1, the gas column density is estimated to be $N_{\mathrm{H_2}} \sim 1 \times 10^{23}$~cm$^{-2}$, which corresponds to a gas number density of $\sim 2 \times 10^{7}$~cm$^{-3}$.

\restartappendixnumbering
\section{Spectral energy distributions} \label{sec_app_sed} 
\begin{figure}[tpbh!]
\begin{center}
\includegraphics[width=11.6cm]{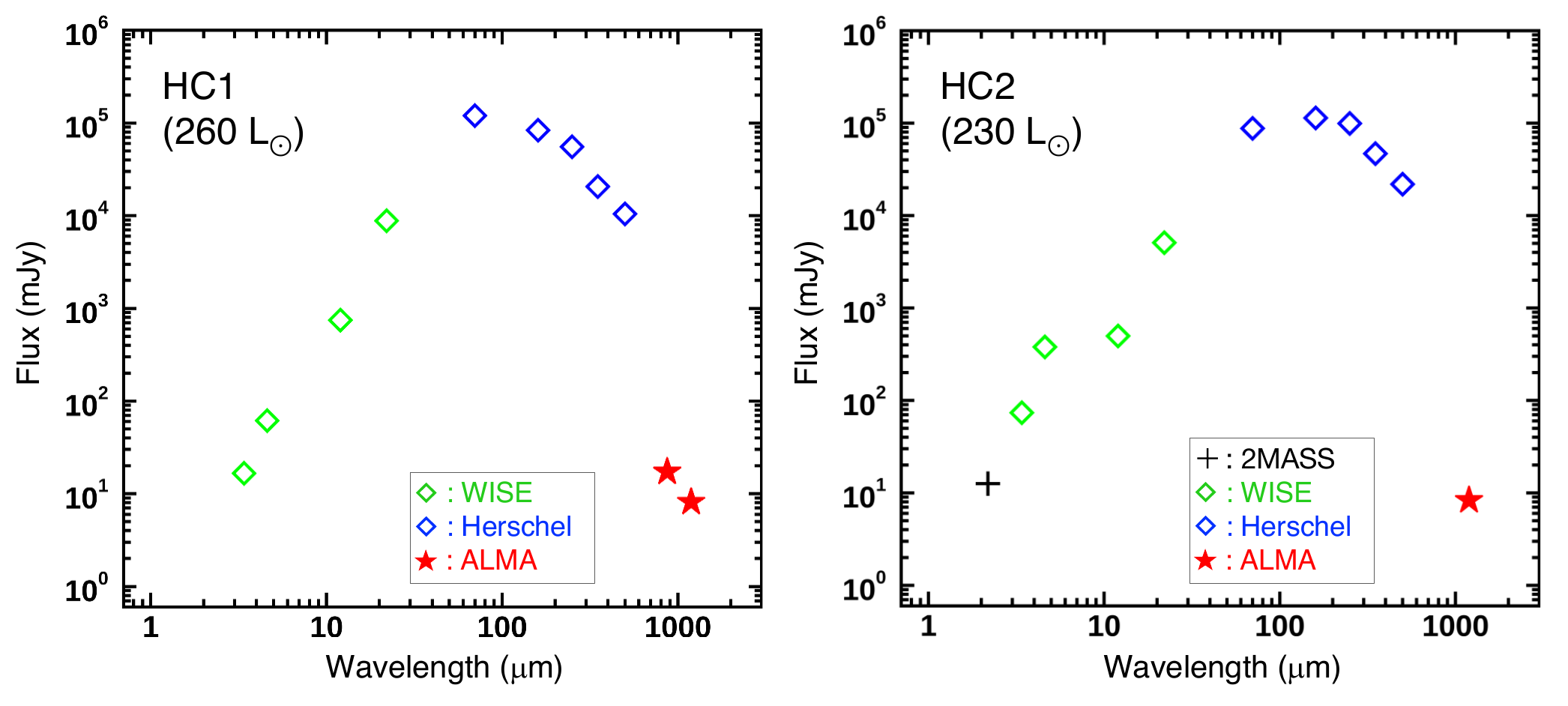}
\caption{
The SEDs of RX1713 HC1 and HC2. 
}
\label{sed}
\end{center}
\end{figure}

Figure~\ref{sed} shows the SEDs of RX1713 HC1 and HC2, which are discussed in Section~\ref{sec_disc_phys}. 
The photometric data are compiled from 
the Two Micron All Sky Survey catalog \citep[][DOI: 10.26131/IRSA2]{Skr06}, 
the AllWISE catalog \citep[][DOI: 10.26131/IRSA1]{Wri10}, and 
the Herschel infrared Galactic Plane Survey (Hi-GAL) catalog \citep[][DOI: 10.26131/IRSA27, 10.26131/IRSA25, 10.26131/IRSA24, 10.26131/IRSA28, 10.26131/IRSA26]{Mol16}, and were obtained from the NASA/IPAC Infrared Science Archive. 
For the Herschel far-infrared data, which are crucial for determining the bolometric luminosity, the angular resolution is coarse compared to the present ALMA observations ($\sim$6\arcsec\ and 11\arcsec\ at 70 and 160~$\mu$m). 
However, the far-infrared emission is detected as point-like sources peaking at the positions of HC1 and HC2, and we therefore consider that the fluxes from these point sources dominate their luminosities. 
A more accurate determination of the luminosities of the regions traced by the present ALMA data will require future far-infrared observations with higher angular resolution. 

\restartappendixnumbering
\section{Lifetime of gas-phase molecules with an enhanced ionization rate} \label{sec_app_networkmodel}
Here we discuss the destruction timescale of gas-phase molecules detected in RX1713 hot cores. 
For this purpose, the gas-phase chemical network is numerically solved utilizing the astrochemical code, Rokko \citep{Fur15}. 
We adopt the chemical network of \citet{Gar13} as our basis, but a large fraction ($\sim$80 \%) of the gas-phase reactions has been replaced with or taken from KIDA2024 \citep{Wak24}. 
Exceptionally, reactions involving O- and N-bearing COMs are still mostly taken from \citet{Gar13}. 
The main destruction pathway of COMs in hot cores has been suggested as reactions with atomic H or the proton transfer from H$_3$O$^+$ followed by the dissociative recombination \citep[e.g.,][]{NM04}. 
For example, proton transfer from H$_3$O$^+$ to CH$_3$OH proceeds as an exothermic process, because of the higher proton affinity of CH$_3$OH compared to H$_2$O.
Protonated methanol can be destroyed by the recombination with electrons, which favors breaking the original structure (e.g., leading to CH$_3$ + OH + H) rather than the formation of methanol (CH$_3$OH + H) \citep{Gep06}. 
This situation is similar for the electron recombination of protonated methyl formate \citep{Ham10}. 
\citet{Taq16} proposed the importance of proton transfer reactions between NH$_3$ and protonated COMs in this context. 
NH$_3$ has the highest proton affinity among the major components of interstellar ices and its proton affinity is higher than that of COMs detected in this work. 
Then, the proton transfer from the protonated COMs to NH$_3$ is exothermic, and can suppress the destruction of the COMs. 
We add relevant proton transfer reactions taken from \citet{GH23} to our network. 
Note that we do not distinguish the ionization by X-rays from that by cosmic-rays for simplicity.

We run a small set of pseudo-time-dependent models, varying in the ionization rate while fixing the H$_2$ gas density at 10$^7$ cm$^{-3}$, gas temperature at 150 K, and the visual extinction at 20 mag.
The initial molecular abundances are given in Table \ref{tab_initab}, referring to the typical interstellar ice compositions \citep{Boo15} and the COM compositions of HC1 listed in Table \ref{tab_N}.
The elemental abundances of the other elements included in our network (He, Si, Fe, Na, Mg, Cl, and P) are taken from \citet{AH99} and they are assumed to be initially in atomic forms.
We consider the range of the ionization rate of \ce{H_2} ($\zeta$) from 10$^{-17}$ s$^{-1}$ to 10$^{-14}$ s$^{-1}$. 

\begin{figure}[bhp!]
\begin{center}
\includegraphics[width=15.0cm]{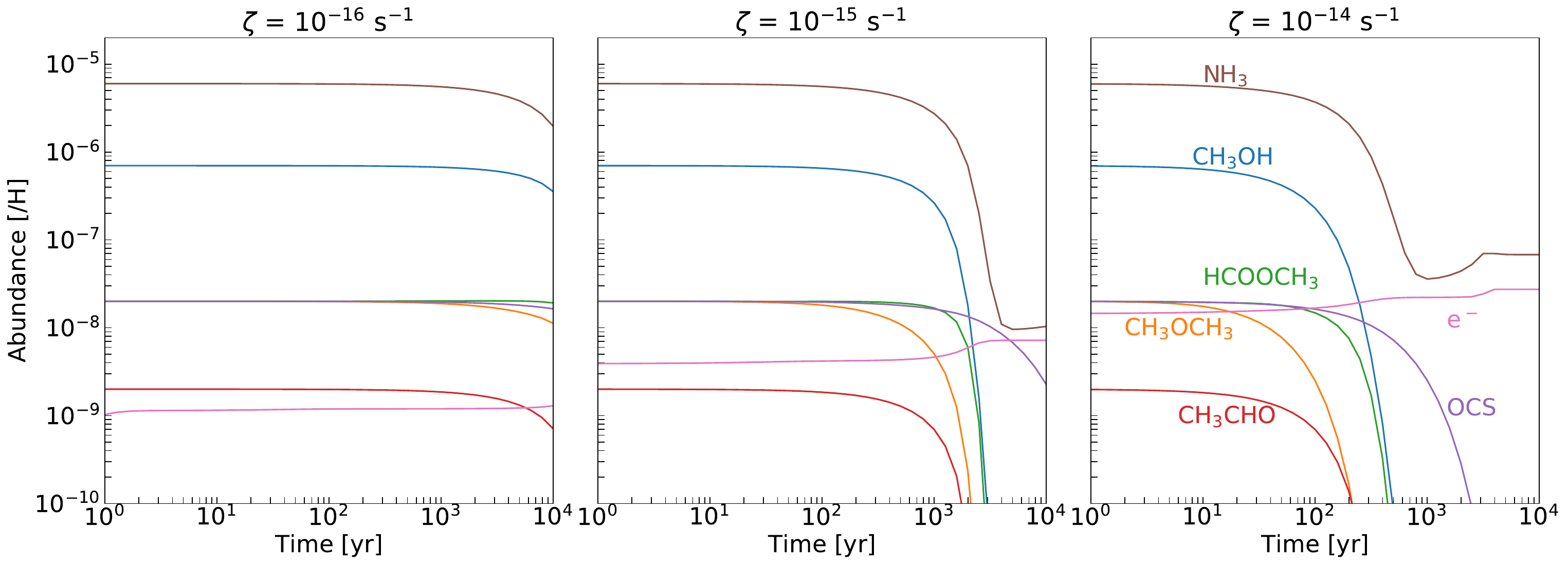}
\caption{Temporal evolution of the abundances of selected gas phase molecules with respect to hydrogen nuclei. The left, middle, and right panels shows the models with the ionization rate of 10$^{-16}$ s$^{-1}$, 10$^{-15}$ s$^{-1}$, 10$^{-14}$ s$^{-1}$, respectively.}
\label{fig_model}
\end{center}
\end{figure}

Figure \ref{fig_model} shows the temporal evolution of the abundances of selected gas phase molecules. 
The lifetime of the gas-phase COMs is inversely proportional to $\zeta$, as the rate-determining step for their destruction is the ionization of H$_2$. 
As long as the ionization rate is lower than 10$^{-15}$ s$^{-1}$, the gas-phase abundances do not change over a 10$^{3}$ yr timescale.

\begin{table}
\caption{Initial molecular abundances with respect to hydrogen nuclei. \label{tab_initab}}
\begin{center}
\scriptsize
\begin{tabular}{cccc}
\hline\hline   
Species &  Abundance &  Species &  Abundance  \\
\hline
\hline
H$_2$  & 0.5  & HCOOCH$_3$ & $2\times10^{-8}$ \\
H$_2$O & 10$^{-4}$ & CH$_3$OCH$_3$ & $2\times10^{-8}$ \\
NH$_3$ & $6\times10^{-6}$ & CH$_3$CHO & $2\times10^{-9}$ \\
N$_2$  & $3\times10^{-5}$ & C$_2$H$_5$OH & $5\times10^{-9}$\\
CO$_2$ & $3\times10^{-5}$ & CH$_3$CN & $3\times10^{-9}$ \\
CO     & $2\times10^{-5}$ & C$_2$H$_5$CN & $5\times10^{-10}$ \\
CH$_3$OH & $7\times10^{-7}$ & OCS & $2\times10^{-8}$ \\
\hline
\end{tabular}
\end{center}
\end{table}

\end{document}